\documentclass[acmtog]{acmart}

\AtBeginDocument{%
  \providecommand\BibTeX{{%
    \normalfont B\kern-0.5em{\scshape i\kern-0.25em b}\kern-0.8em\TeX}}}

\setcopyright{acmcopyright}
\copyrightyear{2021}
\acmYear{2021}

\usepackage{subcaption}
\usepackage{algorithm}
\usepackage{algorithmicx}
\usepackage{xspace}
\usepackage[noend]{algpseudocode}
\usepackage[nounderscore]{syntax}
\usepackage{graphicx}
\usepackage{tikz}
\usepackage{gensymb}
\usepackage{enumitem}
\usepackage[normalem]{ulem}
\usepackage{float}
\usetikzlibrary{matrix,shapes,arrows,fit}

\definecolor{darkgreen}{rgb}{0,0.5,0}
\definecolor{orange}{rgb}{1,0.5,0}
\definecolor{teal}{rgb}{0,0.5,0.5}
\definecolor{darkpurple}{rgb}{0.5, 0, 0.5}
\definecolor{deepskyblue}{rgb}{0.0,0.75,1.0}

\newcommand{\BD}{\mathbf{D}}
\newcommand{\BQ}{\mathbf{Q}}

\DeclareMathOperator{\Tr}{tr}

\begin{document}

\title{Differentiable 3D CAD Programs for Bidirectional Editing}

\author{Dan Cascaval}
\email{cascaval@cs.washington.edu}
\affiliation{%
  \institution{University of Washington}
}

\author{Mira Shalah}
\email{mira@cs.stanford.edu}
\affiliation{%
  \institution{Stanford University}
}

\author{Phillip Quinn}
\email{paquinn@cs.washington.edu}
\affiliation{%
  \institution{University of Washington}
}

\author{Rastislav Bodik}
\email{bodik@cs.washington.edu}
\affiliation{%
  \institution{University of Washington}
}

\author{Maneesh Agrawala}
\email{maneesh@cs.stanford.edu}
\affiliation{%
  \institution{Stanford University}
}

\author{Adriana Schulz}
\email{adriana@cs.washington.edu}
\affiliation{%
  \institution{University of Washington}
}

\begin{abstract}
  Modern CAD tools represent 3D designs not only as geometry, but also as a program composed of geometric operations, each of which depends on a set of parameters. Program representations enable meaningful and controlled shape variations via parameter changes. However, achieving desired modifications solely through parameter editing is challenging when CAD models have not been explicitly authored to expose select degrees of freedom in advance.
  We introduce a novel bidirectional editing system for 3D CAD programs. In addition to editing the CAD program, users can directly manipulate 3D geometry and our system infers parameter updates to keep both representations in sync.
  We formulate inverse edits as a set of constrained optimization objectives, returning plausible updates to program parameters that both match user intent and maintain program validity. Our approach implements an automatically differentiable domain-specific language for CAD programs, providing derivatives for this optimization to be performed quickly on any expressed program.
  Our system enables rapid, interactive exploration of a constrained 3D design space by allowing users to manipulate the program and geometry interchangeably during design iteration. While our approach is not designed to optimize across changes in geometric topology, we show it is expressive and performant enough for users to produce a diverse set of design variants, even when the CAD program contains a large number of parameters.
\end{abstract}




\begin{teaserfigure}
  \includegraphics[width=\linewidth]{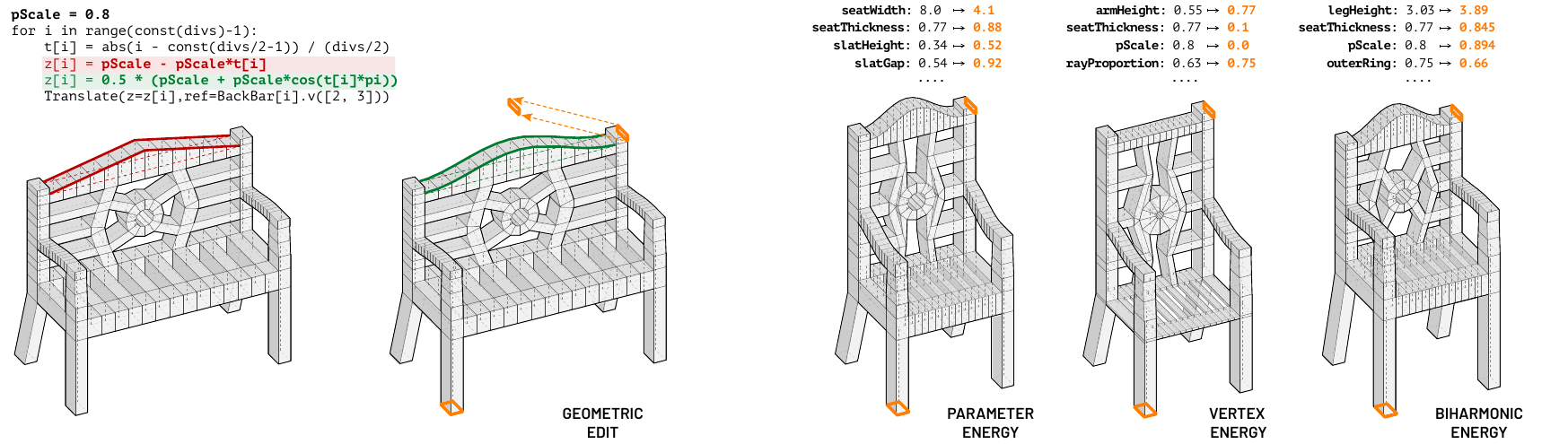}
  \caption{A user bidirectionally edits a model of a bench. They modify the program to change the sharp backrest~(red) to a smooth one~(green). Then, the user intends to rapidly change the bench into a chair. To accomplish this the user performs a geometric edit~(orange), dragging the top part of the leg upwards and in, and also fixing in place vertices on the opposite leg. Our system optimizes program parameters to match the user's edit according to a variety of energy minimization heuristics (Section~\ref{sec:optimization}); returning distinct options for the user to select between. Each option changes between 8 to 12 parameters~(top) from the original program to satisfy the edit.}
  \label{fig:teaser}
\end{teaserfigure}

\newcommand{\hsyntax}[1]{\ensuremath{\mathit{#1}}}

\newcommand\norm[1]{\left\lVert#1\right\rVert}
\newcommand{\T}[2]{\ensuremath{T_{v_{#1}}(#2)}}
\newcommand{\Tx}[1]{$T_{v_{#1}^x}(p)$}
\newcommand{\Ty}[1]{$T_{v_{#1}^y}(p)$}
\newcommand{\Tz}[1]{$T_{v_{#1}^z}(p)$}
\newcommand{\co}[1]{\ensuremath{C({v_{#1}})}}
\newcommand{\proj}{\mathrm{proj}}

\maketitle

\section{Introduction}

Parametric Computer-Aided Design (CAD) models are ubiquitous in engineering. Modern CAD tools (Solidworks, Fusion 360, Onshape, Inventor) represent part geometry using a sequence of parameterized operations, which are executed to produce a 3D shape. Such sequences can be thought of as programs, which offer a form of structured editing: engineers can change operation parameters and the geometry will update without a user having to manually model each component of the output. This is especially powerful when automatically generating repetitive model components such as grids or patterns, and as a module of reuse between different geometric designs. However, manipulating models solely through parameter changes is often cumbersome. Editing a model to achieve a specific design change might require changing many different operation parameters in concert, and real-world manipulations can involve manipulating tens of parameters in a model made up of hundreds of distinct operations.

In contrast, directly manipulating geometry by moving vertices or faces in 3D allows users to specify their desired output without having to translate design intent into a given model's parameter space. The trade-off is that direct manipulation is often imprecise and requires the user to directly model all aspects of a change, which can be both tedious and error-prone, resulting in a model which is inconsistent or physically inaccurate. However, direct manipulation does offer a number of advantages: feedback is immediate, changes are limited to what the user manipulates, and it is clear how to edit any subsection of the model.

Neither editing paradigm is suitable for all manipulations; the right tool for the job depends on the edit itself, which cannot be known in advance. The primary contribution of this work is merging the two paradigms. We extend program-based CAD modeling to allow {\em bidirectional} editing: users can either adjust program operations, or directly manipulate a chosen subset of the output geometry and obtain corresponding operation parameters. As a result, CAD users and tools can leverage whichever representation is more convenient for a given manipulation. To interleave direct and program edits, it is critical that direct manipulations preserve program structure. Accordingly, in this work we focus on resolving direct manipulations in terms of model parameter changes, without changing program structure, and focus on interactive performance to allow for rapid iterations and geometric adjustments.

The fundamental challenge in designing this bidirectional interface is the gap between tens or hundreds of parameters exposed when composing constructive CAD operations into programs, and the handful of degrees of freedom that users may wish to manipulate during a given edit. While prior work has relied on laborious manual specification of these degrees of freedom and their valid ranges~\cite{Schulz:2017,MB:2021:DAGA} our goal is to enable direct editing through optimization over the constructive parameters themselves. When parameter interdependencies and constraints have not been manually exposed by an expert, there tend to be two main problems: ambiguity, as many plausible solutions can match a direct geometric edit, and safety, as many other parameter combinations result in invalid geometry or execution failure.

To address ambiguity, we formulate the inverse editing problem as a series of constrained optimizations over a set of heuristics, and provide a suggestive interface for selecting the desired solution. Interactive optimization times are challenging to achieve due to the dimensionality of the parameter space and the need to solve many optimization problems in parallel, one for each heuristic. To solve this optimization quickly, we developed a Domain-Specific Language (DSL) for CAD models that is automatically differentiable, aiming to express many common geometric operations while being able to provide derivatives for any expressible program.

Since naively calling automatically differentiated CAD kernels is impractical in an editing context due to performance concerns~\cite{Banovic:2019}, our solution evaluates model code to produce a computation graph that does not refer to the underlying CAD kernel. This in turn enables optimization on the generated code without differentiating the underlying engine, or executing any of its extra overheads while optimizing (Section~\ref{sec:implementation}). A language-based approach also allows us guarantee safety: we choose operators and implement operation semantics such that discontinuities and execution failures are prevented by definition.

In summary, our primary contribution is {\bf a system for bidirectional editing of 3D CAD models with interactive performance}. We demonstrate how such a system can be constructed to handle real-world models by using a domain-specific language, and describe our solutions to the specific engineering concerns that must be addressed to make it usable in practice. We evaluate our system's editing capability and performance on a series of case study models, demonstrating its efficacy at finding viable parameter combinations to satisfy edits on large models.

\begin{figure}
  \includegraphics[width=\linewidth]{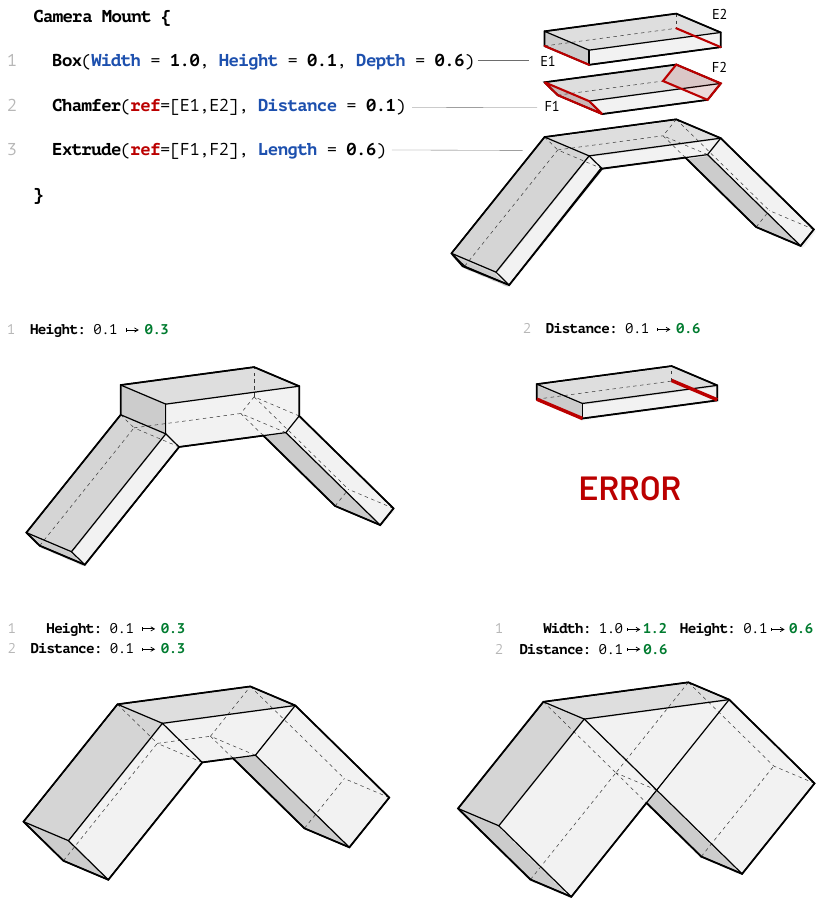}
  \caption{Attempting to thicken the camera mount uniformly by changing a single parameter (middle left) loses the overall structure of the model as the thickness of the base differs from the thickness of the legs. If we instead modify the chamfer radius (middle right) the program does even run to completion, as it would result in non-manifold geometry. To perform thickening, our tool automatically updates multiple parameters in concert to achieve the desired properties (bottom) while staying within the execution constraints.
  }
  \label{fig:errors}
\end{figure}

\section{Background and Related Work}
Editing program-based models is of interest to the Computer Graphics, CAD, and Programming Languages communities. Each of these fields have contributed their own ideas and work on the problem, and our work draws inspiration from all three.

\paragraph{Editing CAD Models}
Many modern CAD tools (Solidworks, Fusion 360, Inventor, Onshape) use programs in the form of a history of operations that are executed to construct a model. These are called {\em history-based} systems. To address the challenges of editing such CAD programs, some of these systems incorporate direct edits (e.g. moving a face) by appending operations to the history. Adding operations in this manner is limited to predefined manipulations, and changes the {\em structure} of the CAD program and types of shape variations it can produce -- they may break existing symmetries, and force users to consider the appended operations when making changes to the model.

Another approach is to ignore the CAD program entirely and directly edit the geometry, without preserving any notion of history (e.g. SpaceClaim, KeyCreator, and Rhino). This {\em history-free} approach loses the benefits provided by the program representation \cite{Whatsth19:online,Thefaile61:online}. Siemen's Synchronous Technology takes a hybrid approach and integrate history-free geometric editing on top of a program/history~\cite{synctechinfo}.
However, by partially disregarding the history, the CAD model can no longer be expressed or analyzed as a single program that defines semantically meaningful shape variations~\cite{SolvedRe10:online,Parametr28:online}. In practice, this means that after a model is manipulated directly, existing constraints (such as symmetry and spacing) may be violated, and unexpected breakages may occur when the part is later updated parametrically and a manipulation cannot be reconciled.

Unlike these methods, our {\em bidirectional} editing approach preserves the history as an explicit program and allows the user to seamlessly switch between editing the program structure and parameters, and directly editing its geometry.

\paragraph{Program Editing} While prior work in programming languages has investigated solutions to bi-directional program editing~\cite{Hu:2011:BX} in domains like layout design and SVG, our method differs in how we address {\em ambiguity} that results from a user manipulation.

Such methods eliminate ambiguities by either restricting user interactions~\cite{Hottelier:14:PBM} or pre-defining which parameters a manipulation will affect~\cite{Chugh:2016:1,hempel2019sketch}. Our method instead aims to expose more to the user by allowing arbitrary manipulations and returning a variety of different solutions in the case of ambiguity.

There has also been increased effort to reconstruct CAD programs from existing geometry~\cite{du2018inversecsg,nandi2018functional,willis2020fusion} as well as re-writing CAD programs to expose meaningful parameters~\cite{nandi2020synthesizing}. While none of these methods propose solutions to interactive manipulation, they can be modified to generate and refine a CAD program suitable for input to our system, widening the potential application of bidirectional editing beyond human-written programs.

\paragraph{Procedural Model Editing} Shape-generating programs are known in the graphics community as {\em procedural models}~\cite{Wonka:2003:Architecture,Muller:2006:Procedural,Lipp:Visual:2008,Berndt:2005:GML}. Controlling procedural models, however, is a notoriously challenging task~\cite{smelik2014survey}. Many approaches on inverse procedural modeling target program generation from input specifications~\cite{talton2011metropolis,vanegas2012inverse,guo2020inverse}, but do not allow direct editing. While there is some work on local editing from program analysis~\cite{jesus2018generalized,Lipp2019}, none of these methods allow users to control geometric edits directly while maintaining a working program.

Of particular note is the DAG Amendment method described by \citet{MB:2021:DAGA}, which also allows inverse control of procedural models generated by a DAG of operations, using a brush interaction to address edit ambiguity. This work handles operations that can change the output mesh topology as long as an unambiguous mapping from output to input UV coordinates is maintained. Such generality is achieved by performing differentiation using finite differences, which is performant for models that have been designed to expose a small set of degrees of freedom in advance, and handles a wide variety of operations for which it is unclear how to differentiate automatically.

In contrast, our work pursues the opposite tradeoff: we do not handle operations that can result in topological changes, and instead we apply fully automatic differentiation, allowing our approach to scale up to models of tens or hundreds of parameters where finite differences would be prohibitively expensive. This efficiency allows us to perform a full non-linear optimization interactively when resolving edits. As a result, our work can handle real-world CAD models where parameters have not been manually tuned for editing, models where the valid ranges for parameters vary as functions of other parameters, and edit specifications where some model geometry can be fixed in place for an edit duration while multiple other parts are changed.

\paragraph{Structure-Driven Geometry Editing} There is a large body of work on structure-preserving geometric edits using an analyze-and-edit approach to constrain the shape without any underlying program \cite{mitra2014structure}. Previous work has analyzed salient features \cite{kraevoy2008non}, feature curves \cite{Gal:2009:IWIRES}, relations between shape parts \cite{Zheng:2011:Controllers}, replicated patterns \cite{bokeloh2011pattern,bokeloh2012algebraic}, and manufacturability \cite{schulz2014design} as methods of describing and preserving shape structure.

Our work builds upon these ideas to define inverse editing as an optimization. However, unlike previous approaches, our method extends geometric analysis with constraints and objectives encoded through a program construction. We show how encoding programs in our differentiable DSL can produce improved results compared to applying traditional geometric optimization followed by parameter synchronization (see Figure~\ref{fig:biharmonic}).

Furthermore, instead of a single optimization, our system allows users to navigate a gallery of results that are generated by prioritizing different geometric properties in the optimization goals.

\paragraph{Differentiable Programming} Recent years have seen an explosion in the application of differentiable programming techniques in Machine Learning, particularly through automatic differentiation~\cite{griewank2008derivatives}. Packages such as PyTorch~\cite{NEURIPS2019_PY} provide an interface by which a computation graph is automatically constructed by executing user code, and derivative functions are automatically derived from this graph. Others such as JAX~\cite{frostig2018compiling} and XLA~\cite{tfXLA} go a step further, and enable compilation of user functions and their automatic derivatives from high-level dynamic languages into native code for fast optimization and training. Our work takes a similar strategy to these tools, tailored towards the specific optimization problems in our domain.
Existing work using automatic differentiation in computer graphics applications focuses on formulating or recreating traditional graphics pipelines (e.g. rasterization, rendering) as differentiable functions~\cite{li2020differentiable,nimier2019mitsuba}, demonstrating applications that arise from operating directly on structured representations such as vectors or scene primitives, at a higher level than pixel data \cite{reddy2020discovering}. Other work performs finite-element optimization directly on parametric CAD models~\cite{hafner2019x} by providing differentiable physical simulations. Our work extends both of these lines of work, providing a language of composed of differentiable operators that can express a wide range of CAD models, and derivative information for any program in the language. As a result we enable optimization directly on a the structured program representation during the model design and editing loop.

\begin{figure}
  \centering
  \includegraphics[width=\linewidth]{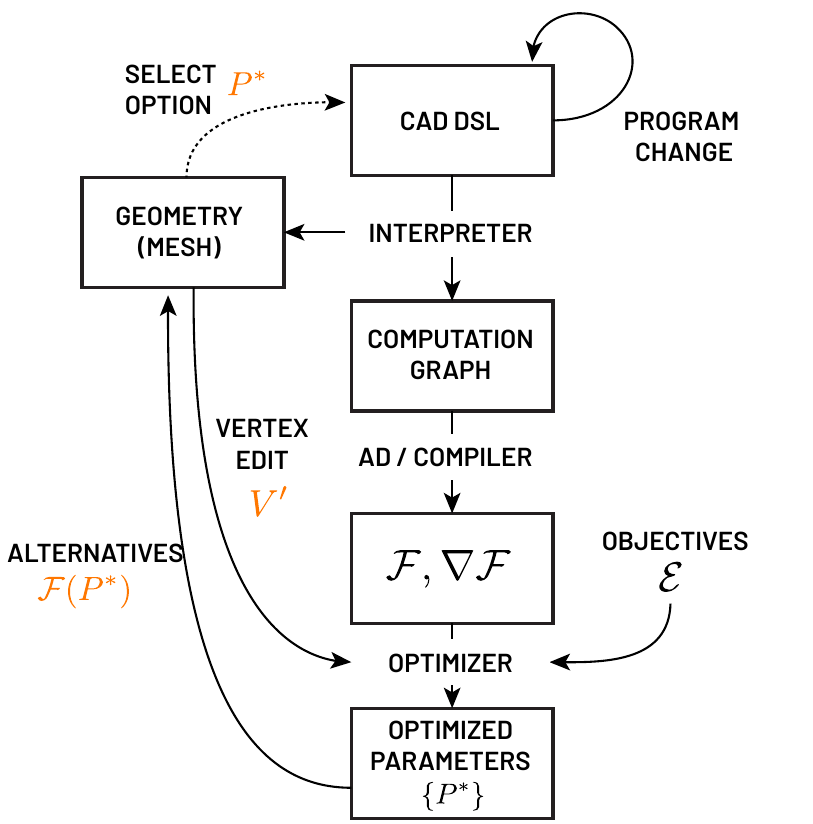}
  \caption{{\bf Pipeline.} A program is interpreted to create mesh geometry and a corresponding computation graph. We extract derivatives from this graph and compile a function $\mathcal{F}$ from parameters to vertices and its derivative for use in optimization with several different energy functions. Users can then edit the mesh, and receive updated sets of program parameters $P^*$ corresponding to each energy. Parameters are then used to compute updated mesh vertex positions $\mathcal{F}(P^*)$ matching the user edit. Users choose and option and can continue editing geometrically, as well as update the program with the found parameters.}
  \label{fig:pipeline}
\end{figure}

\section{Interactive Editing System}

Our goal is to enable interactive bidirectional editing of a CAD program and the 3D geometry it generates. When a user edits the program, our system interprets the edited program to generate the corresponding updated geometry. When the user directly edits the geometry by manipulating vertices, edges, or faces, our system finds sets of updated parameters such that the program, when executed with these parameters, produces geometry where the manipulated elements are as close as possible to the user's edit.

To accomplish this, we
(1) specify 3D models as programs in a CAD language designed to be differentiable;
(2) interpret this language to generate both geometry as well as a computation graph representation of the program, which is then automatically differentiated and compiled into native code;
(3) optimize over program parameters using a set of heuristic objectives that take into account both program and geometric information, serving as proxies for properties ranging from appearance to physical performance. This pipeline is illustrated in Figure~\ref{fig:pipeline}.
\subsection{Language Design}
\label{sec:lang}

Our language makes a few key design choices motivated by the domain and specific problem requirements of bidirectional editing.

\paragraph{Explicit Geometric Representation} It is common in the CAD domain for users to reference features of previously constructed geometry, such as edges or faces. This is often necessary in order to concisely describe complex CAD models with a small number of parameters. In our system, we use a polygonal mesh representation, and users can use vertex positions as parameters in subsequent operations, and reference their values when defining constraints. Relying on an explicit mesh representation makes it easier to define a wide variety of optimization objectives that refer to explicit features of output geometry, such as minimizing the distance to vertices manipulated by a user or minimizing the deformation from the starting mesh.

\paragraph{Static Model Topology} However, such an explicit representation comes at a cost. Since we allow user programs to refer directly to output vertices by index, in order to provide differentiability during inverse editing, we must guarantee that a given operation returns the same number of vertices for all parameter combinations -- if this number were allowed to change during optimization, it would be unclear which vertices are being referred to by an index reference. The same logic applies to the order of an operation's returned vertices. Both of these cases are discrete discontinuities in the model, and it is not clear how to optimize such a program without differentiating repeatedly during each optimization step. We therefore only allow topological changes to be performed on the program, and not inferred through geometric manipulations.

Our DSL can include any operation which creates, modifies, or removes a statically known amount of vertices and faces. Positions of vertices that are created or modified must be differentiable functions of parameters and prior vertex positions. We guarantee that any point in the parameter space of any program in our language is differentiable. Degenerate or zero-size mesh elements do not pose any issue in this framework, and are thus handled robustly. Working within the limitations of this static topology allows us to compile energy functions ahead of time, drastically increasing performance.

\paragraph{Constraints} We also allow direct specification of constraints to further limit the parameter space considered during optimization. Constraints are imposed automatically for some operations, and we expose a mechanism for users to further impose constraints. These constraints are treated as hard requirements, and any result returned by our optimization necessarily respects them. Constraints satisfy three key goals:

Firstly, we would like to avoid parameter changes that lead to non-manifold geometry, which would cause objective functions that are defined on properties such as volume to return inaccurate results. Some constraints to prevent this are automatically generated by the interpreter as part of operation definitions, as we describe in Section~\ref{sec:tracing}. While not all non-manifold cases can be prevented, applying the automatically-generated constraints prevents a large class of undesirable parameter combinations by default.

Secondly, users can apply constraints in order to refine the allowable variations in the program; by eliminating self-intersections or other undesirable model variations if these come up in practice.

Finally, using constraints, users can impose additional design objectives such as maintaining relative proportions, aspect ratios, or distances between mesh elements; without changing the structure of their program to enforce these properties by construction.

\newcommand{\F}[0]{\mathcal{F}}

\subsection{Interpretation and Differentiation}
\label{sec:tracing}

\begin{figure*}
    \centering
    \includegraphics[width=\linewidth]{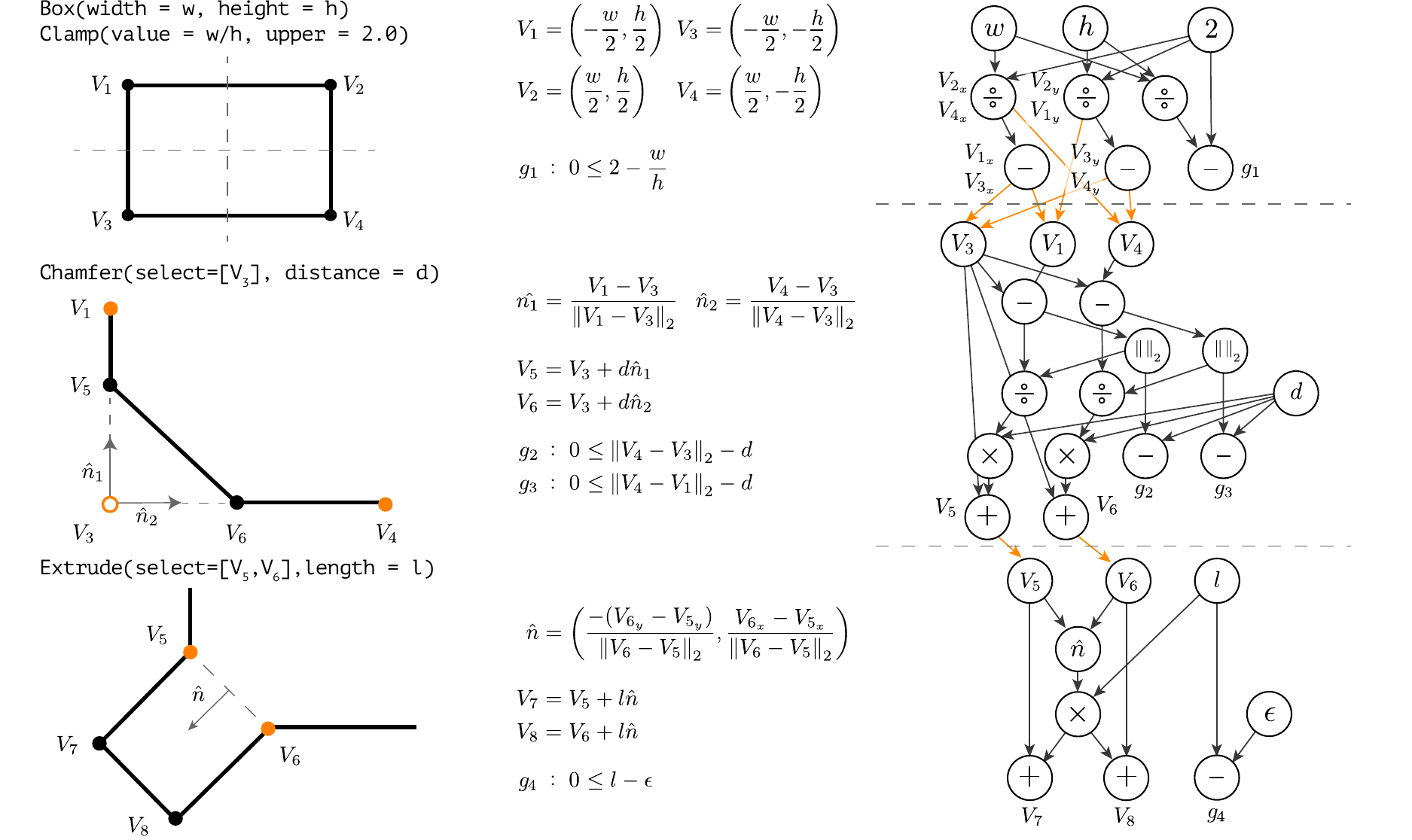}
    \caption{
        To construct the computation graph G for this 2D version of our camera mount program (left), we  trace the operations in our DSL (e.g. {\tt Box}, {\tt Clamp}, {\tt Chamfer}) to generate expressions that create and modify vertex positions (e.g. $V_1$, $V_2$, ...) based on the operation parameters (middle). We encode these expressions in the computation graph $G$ (right). Each operation {\em references} previously constructed geometry (orange) and imposes constraints (e.g. $g_1$, $g_2$).
        We note that, for concision, the nodes in the {\tt Chamfer} and {\tt Extrude} operations are operating on vertex values, rather than scalars. They can be treated as pairwise operations on the constituent elements.}
    \label{fig:tracing}
\end{figure*}

\paragraph{Computation Graph} When executing our CAD DSL program to generate output geometry, our algorithm operates as an interpreter. During the execution of each program operation, in addition to computing the result of the operation, our interpreter also keeps track of the individual computations that are performed, maintaining a directed computation graph $G$ (Figure~\ref{fig:tracing}).

Each node in $G$ is either a constant, a variable, or a differentiable operator $f$, (e.g. {\tt +}, {\tt *}, {\tt sin}, {\tt pow}). Each node tracks the temporary value resulting from the execution of its operator, and each graph edge represents a {\em use} of this value. A vertex coordinate is fully encoded by marking a node in the graph as representing the computation for that coordinate.

Each CAD operation in our program (e.g. {\tt Box}, {\tt Chamfer}, {\tt Extrude}) adds nodes to this graph, often directly referencing the values of vertices output by previous operations as the basis for their computation, as depicted in Figure~\ref{fig:tracing}. Each operation also modifies the markers corresponding to vertices, by either adding additional markers for new vertices, moving markers to new nodes to represent vertex transformations, or removing markers (e.g. $V_3$ of {\tt Chamfer} in Figure~\ref{fig:tracing}) to indicate that, while a computation may still be present and referred to, it no longer corresponds to a vertex.

Note that our method only tracks vertex positions, and does not deal with aspects of mesh topology such as edges and faces. These are created and tracked by an underlying CAD system. This process is described in more detail in Section~\ref{sec:implementation}. For the purpose of inverse editing this topology is static, and we can fully describe parameter updates by adjusting vertex positions.

\paragraph{Enforcing Constraints} Along with computing and tracing vertex positions, for certain operations our interpreter automatically introduces constraints on the parameters as part of the operation definition. For example, the {\em length} parameter of an {\tt Extrude} operation is constrained to always remain larger than some positive epsilon, ensuring geometry is extruded outwards along a face's normal. Accordingly, our interpreter encodes the constraint as the value $length - \epsilon$, which is added to the graph, and marked as a constraint in a fashion identical to vertex coordinates. Similar constraints are generated for {\tt Chamfer}, preventing the radius from exceeding the length of either edge being chamfered ($g_2, g_3$ in Figure~\ref{fig:tracing}).

During optimization, we require nodes marked as constraints to remain positive. As a result, this expression implicitly represents a constraint of the form $g(P_1, ...,~P_k) \geq 0$. This can refer to vertex coordinates, since every graph node is ultimately a function of the parameters.

In addition to automatically-applied constraints, we expose a {\tt Clamp} operation to the program author, which can be used to set upper and lower bounds on the allowed values of an arbitrary parameter expression, for example, bounding the maximum aspect ratio for the rectangle shown in Figure~\ref{fig:tracing}, or bounding the range of a vertex coordinate. When executing ${\texttt{Clamp}}(\alpha,f(P_1,~...,~P_k),\beta)$, we generate constraints $f(P_1,~...,~P_k) - \alpha \geq 0$ and $\beta - f(P_1,~...,~P_k) \geq 0$, with $f$ representing the parameter expression whose value we are bounding, and $\alpha$, $\beta$ representing expressions for lower and upper bounds on $f$, respectively.

\paragraph{Differentiation and Compilation} $G$ implicitly represents a function~$\F$ from our program parameters $P$ to the positions of output vertices~$V$. We can evaluate $\F$ by updating the values of the variable nodes, traversing $G$ to recompute the values for all nodes whose values have been invalidated as a result of a change in an upstream node, and then visiting all of the marked nodes to extract the updated vertex coordinates. Once $G$ is constructed, we extract the gradient $\nabla \F$ through reverse-mode automatic differentiation \cite{margossian19autodiff}, giving us the gradient vector of program parameters with respect to each vertex coordinate. We can extract the parameter gradient with respect to each constraint as well, which is critical for performant and stable optimization.

As a final step in our interpretation pipeline, we compile $\F$, $\nabla \F$, the constraint functions, and their respective gradients to an explicit native code representation, which is passed off to an optimizing compiler. As a result, we can execute $\F$ and $\nabla\F$ several orders of magnitude more quickly than by traversing $G$, as all of the overhead of the graph representation has been compiled away. Compilation is necessary to maintain interactivity during program editing and synchronization, which evaluates $\F$ composed with each objective function (Section~\ref{sec:optimization}) repeatedly. Compilation can take significantly more time than interpretation, so instead of compiling after every program change, we implement this as a separate step that a user performs in order to begin synchronizing the model with geometric edits. Consecutive synchronization steps can re-use a compiled representation until the program is changed, and users only ever need to compile before making a geometric edit immediately after a program edit.
\subsection{Optimization}
\label{sec:optimization}

When the uses directly edit geometry, they adjust the positions of a set of vertices $V'{\subset}V$. Unless they manipulate all of the vertices, this is a {\em partial} specification: the user is providing no information about the vertices $v_i \notin V'$. There are two primary challenges in solving for the program parameters $P$, given this edit:
\begin{itemize}[leftmargin=*]
    \item The edit may not be {\em feasible}, if there does not exist any $P$ such that the specification above is met while also satisfying the program constraints $g_1, ... ~,~g_k$.
    \item There may be many sets of parameters $P$ that satisfy the specification and it is {\em ambiguous} which one should be returned.
\end{itemize}
Our solution casts the problem as a constrained optimization. We address feasibility by formulating how close a set of parameters $P$ is to satisfying a vertex edit as a continuous energy:

\begin{equation}
    E_{\text{edit}}(P) =\sum_{v_i\in V'}\norm{V_i(P)-\mathbf{\bar{v}_i}}_2^2
    \label{equation:editloss}
\end{equation}

where $\mathbf{\bar{v}_i}\in\mathbb{R}^3$ denotes the spatial position of vertex $v_i$ after the user's edit and $V_i(P)$ denotes the position of $v_i$ when the program is executed with parameters $P$.

However, the challenge of ambiguity remains. Optimization offers a solution here too: we formulate a number of heuristic objectives that can be combined with the edit energy to preserve characteristics of the model while adhering to the user manipulation. While our heuristics must be defined on the parameters $P$, by executing and differentiating $\F(P)$ we can define heuristic notions of change over vertex positions and optimize them through parameter changes. This ability to define change over geometry allows us to use any of a large set of well understood heuristics  with our system as long as they are differentiable.

We describe a few heuristics included in our system and the characteristics they are designed to preserve. As not all objectives will be useful on all classes of models, we allow users to select a subset of the following objectives to apply. Additionally, we demonstrate that while the heuristics do not comprise a fully automatic solution to the ambiguity issue, they are capable of providing {\em diverse} solutions~(Section~\ref{sec:results}), allowing users to explore the solution space.

\subsubsection{Geometric Objectives}
\label{sec:geometricobjectives}

One category of heuristic is based on the assumption that the geometric features that are not manipulated should change as little as possible. One method for minimizing geometric change is to minimize the change in positions of the vertices that were not manipulated by the user, namely, $v_i \notin V'$. The corresponding energy term for this minimization is
\begin{gather}
    E_{\text{vtx}}(P) = \sum_{v_i {\in} V\setminus V'} W_{\text{vtx}}(v_i)||V_i(P)-V_i(P_0)||_1 \\
    W_{\text{vtx}}(v_i) = \frac{\Delta(v_i,V')}{\sum_{u_i {\in} V\setminus V'} \Delta(u_i,V')}
\end{gather}
where $P_0$ denotes the original program parameters prior to optimization.  We weight the contribution of each vertex to the energy by a {\em localization} term $W_{\text{vtx}}$. $\Delta(v_i,V')$ is defined as the shortest geodesic distance between $v_i$ and the unmanipulated position of any vertex in~$V'$, encoding the notion that users are more likely to want changes closer to the vertices they manipulated; as opposed to changes in geometry further away. We use an~$L_1$ norm to prioritize sparsity, preferring to move a few vertex positions significantly rather than moving many vertices a little, as the~$L_2$ norm would promote.

Similarly, we can aim to preserve the lengths of edges in the mesh $(V,E)$ not modified by the user, that is, the edges $e_{ij} \notin E'$, where $E' {=} \{ e_{ij} {\in} E : v_i {\in} V' \land v_j {\in} V' \}$. The corresponding energy is:

\begin{gather}
    E_{\text{edg}}(P) =
    \sum_{e_{ij}\notin E'} W_{\text{edg}}(e_{ij})E_{\text{edg}}(e_{ij})  \\
    E_{\text{edg}}(e_{ij}) = \left( \norm{V_i(P)-V_j(P)}_2-\norm{V_i(P_0)-V_j(P_0)}_2\right)^2 \\
    W_{\text{edg}}(e_{ij}) = \max(W_{\text{vtx}}(v_i),W_{\text{vtx}}(v_j))
\end{gather}

As in our vertex energy, we include a localization weight~$W_{\text{edg}}$ for each edge which is the maximum vertex weight of either of the edge's endpoints.

\subsubsection{Deformation Objectives}

A common approach in interactive surface modeling is to minimize {\em geometric deformation} for all local areas when the user manipulates a few vertices of a given mesh, potentially moving vertices far away from the user's edit to achieve a smoother overall transform. We demonstrate the application of two well-studied heuristics from the surface modeling literature to accomplish this for our system. The first heuristic is that the deformation should be as smooth as possible, which is achieved by minimizing the bi-harmonic energy~\cite{sorkine2004laplacian}.
The second heuristic targets a deformation that is as-rigid-as-possible (ARAP), i.e., it penalizes non-rigid transformations of the model~\cite{sorkine2007rigid}. ARAP is a physics-inspired energy that preserves volume and local structure.

\begin{figure}
    \centering
    \includegraphics[width=\linewidth]{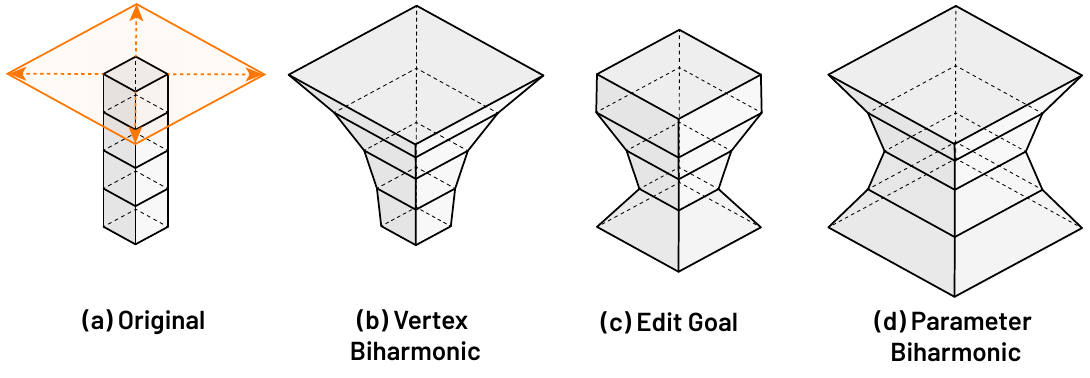}
    \caption{For the model in~(a), the $XY$ coordinates of the vertices at the top and bottom loops are controlled by the same parameter. When the top vertices are edited (orange), (b) shows the result of minimizing the bi-harmonic energy on the vertex positions, disregarding the program. (c) is the result of using the vertex positions in~(b) as $V'$ to minimize our edit objective (Equation \ref{equation:editloss}), whereas optimizing directly on the parameters produces the desired smooth result~(d).}
    \label{fig:biharmonic}
\end{figure}

A naive approach to minimizing these energies might be to perform the deformation directly on the geometry, and then treat the resulting vertex positions as $V'$ for our edit objective. However, this approach bypasses the constraints of the program and leads to non-smooth results compared to optimizing directly on the parameters, as can be seen in Figure~\ref{fig:biharmonic}.

Instead, we optimize the bi-harmonic and ARAP energies directly on program parameters, which define our degrees of freedom for the deformation, instead of on vertex positions. For both deformation energies we first triangulate the original mesh. This only changes the connectivity, and the vertices can still be represented as functions of the parameters by invoking $\F(P)$. The bi-harmonic energy is then formulated as:
\begin{gather}
    E_{bh}(P) = \Tr(\BD(P)^T \BQ \BD(P))
    \label{eq:bh}
\end{gather}
where $\BQ$ is the bi-laplacian of triangulated mesh and $D(P)$ is the displacement vector defined as $\BD(P) = \F(P)-\F(P_0)$.

The ARAP energy can be similarly expressed in terms of the program parameters:
\begin{gather}
    E_{ARAP}(P) =
    \sum_{i=1}^{n} w_i \sum_{j\in \mathcal{N}(i)} w_{ij} \left\Vert(V_i(P){-}V_j(P)){-} R_i(V_i(P_0){-}V_j(P_0))\right\Vert^2
    \label{eq:arap}
\end{gather}

where $\mathcal{N}(i)$ is the one-ring  neighborhood of  vertex~$v_i$ given the triangulated mesh, $R_i$ is the approximate rigid transformation in~$\mathcal{N}(i)$, and~$w_{ij} = \frac{1}{2}(\cot \alpha_{ij}+\cot \beta_{ij})$, where~$\alpha_{ij}$, $\beta_{ij}$ are the angles opposite of the mesh edge $(v_i, v_j)$, and $w_i = 1$.

Since the bi-harmonic function can be analytically differentiated with respect to the deformation $D$, we can further speed up our optimization by evaluating this analytic derivative using a forward automatic differentiation pass for each parameter to create the Jacobian matrix $\frac{\partial \BD}{\partial p_i}$. The implementation of ARAP is similar, although with the addition of the local optimization and iteration~\cite{sorkine2007rigid}. See Appendix~\ref{app:1} for more details.

\subsubsection{Program-driven Objectives}
Optimization towards geometric or deformation objectives produces meshes that geometrically resemble the initial mesh. However, just as with vertex positions and edge lengths, parameter values can also be a proxy for similarity, especially when parameters correspond to semantic attributes of the model, like length, width, or radii. We minimize change in parameter values according to an $L_1$ energy, effectively focusing changes on a minimal subset of program parameters. This is given by
\begin{gather}
    E_{\text{par}}(P) = \sum_{P_i \in P} W_{\text{par}}(P_i)\norm{p_i - p_{0_i}}_1 \\
    W_{\text{par}}(P_i) = \left(1 - \frac{|V_{P_i} \cap V'|}{|V_{P_i}|}\right)^2
\end{gather}
where $p_{0_i}$ is the initial value of parameter $P_i$ at the start of the optimization, and $V_{P_i}$ is the set of vertices whose coordinates in the computation graph $G$ contain $P_i$ as an ancestor. Our localization term $W_{\text{par}}(P_i)$ focuses change in the {\em parameter} space on parameters that correspond more closely with the vertices affected by a manipulation, even if that change might correspond to changes in many vertices or edge lengths. This objective encodes the hypothesis that a small subset of parameters in a CAD program can correspond to a higher-level design criterion, and focusing variation to just these parameters can result in variation along that design criterion, regardless of the underlying geometry. For example, this objective promotes uniformly widening the elements in a model according to a thickness parameter, even if this results in geometric changes that would be penalized by the purely geometric objectives discussed in Section~\ref{sec:geometricobjectives}. An example of this can be seen in Figure~\ref{fig:gallery}b, where the parameter energy results in thickening the entire dresser frame.

An outstanding issue with this approach is that one parameter might be much more sensitive than another. Moreover, there is no clear comparison between a parameter that controls rotation angle and one which controls a length -- the objective will penalize each change equally, regardless of its effect on the output. In our experiments, we were able to achieve reasonable results with the energy as-is, mainly due to induced sparsity and localization. We propose a more in-depth treatment of parameter normalization as part of future work.

\subsubsection{Performance Objectives}
We also present objectives that aim to proxy some aspect of physical performance. For example, preserving volume can be used to maintain a fabrication objective such as material usage:
\begin{equation}
    E_{vol}(P) = \left(Vol(P) - Vol(P_0)\right)^2
\end{equation}
where $Vol(P)$ represents the volume of the output mesh at parameters $P$. Another example is preserving the center of mass, which may be essential for stability:
\begin{equation}
    E_{\text{cm}}(P)= \norm{COM(P) - COM(P_0)}_2 + K_vE_{\text{vtx}}(P)
    \label{equation:COM}
\end{equation}
where $COM$ is the center of mass given model parameters, and $K_v$ is a weighting term. Such performance objectives are often inadequate for the goal of resolving ambiguity -- while they reduce the result space by maintaining a property, there may still be many valid parameter combinations that satisfy the edit. As a result we find it useful to combine these terms with the other geometric energies, particularly the localized vertex objective, using $K_v$ as 0.01 in our implementation.

\subsubsection{Composed Objective.}

Each of the above objectives is then individually combined with our edit objective (Equation~\ref{equation:editloss}) to create
\begin{equation}
    \mathcal{E}_{\text{obj}} = E_{\text{edit}} + \gamma_{\text{obj}}E_{\text{obj}}
\end{equation}
where $\gamma_{\text{obj}}$ represents a weighting term for how strongly the optimization should aim to preserve the property represented by $E_{\text{obj}}$ as opposed to lowering the weight of the edit energy $E_{\text{edit}}$. These two objectives are often conflicting for geometric objectives, where no parameter combination $P$ that minimizes $E_{\text{edit}}$ also minimizes $E_{\text{obj}}$; causing our overall minimization of $\mathcal{E}_{\text{obj}}$ to deviate from satisfying the user's edit. Exposing control of $\gamma_{\text{obj}}$  allows the user to tune how strongly to enforce their edit against an objective. Each $\mathcal{E}_{\text{obj}}$ is treated as a single objective for our optimization step, and the user is given a choice of which objectives to run for a given model. We note that it is also possible to combine multiple different $E_{\text{obj}}$ together with $E_{\text{edit}}$ to form new ones, as done in Equation~\ref{equation:COM}.

\begin{table}
  \resizebox{0.48\textwidth}{!}{%
    \begin{tabular}{lrrrrr}
      \textbf{Model} & \textbf{Size: Ops / Verts} & \textbf{Interpret} & \textbf{Compile} & \textbf{Sync (10)} & \textbf{Sync (More)} \\
      \hline
      Mount          & 221 / 20                   & 0.02s              & 0.27s            & 0.08s              & --                   \\
      Dresser        & 1593 / 468                 & 0.39s              & 2.06s            & 0.75s              & 1.58s                \\
      Chair          & 2534 / 220                 & 0.28s              & 2.40s            & 0.69s              & 0.88s                \\
      Castle         & 11435 / 1481               & 1.22s              & 9.99s            & 1.89s              & 6.14s                \\
      Slipper        & 14835 / 269                & 0.45s              & 3.33s            & 0.73s              & 1.22s                \\
      Chandelier     & 19700 / 1554               & 1.35s              & 16.26s           & 1.02s              & 4.10s                \\
      Lamp           & 29033 / 520                & 0.55s              & 14.87s           & 4.66s              & 18.34s               \\
    \end{tabular}
  }
  \vspace{0.05in}
  \label{tab:perf-table}
  \caption{{\bf Performance metrics.} Operations represent arithmetic operations as tracked by our computation graph. Sync represents the total combined time of optimizing 6 objectives: $E_{\text{edit}}$, $E_{\text{vtx}}$, $E_{\text{edg}}$, $E_{\text{par}}$, $E_{\text{bh}}$ , and $E_{\text{vol}}$. Note that synchronization time can vary greatly depending on the number and position of vertices selected -- we present the mean of optimizing edits of 10 vertices over 10 random edits each; to give a sense of relative performance. Sync (More) contains edits that select more than 50\% of the model's vertices, exceeding the amount found in typical edits.}
\end{table}

\section{Implementation}
\label{sec:implementation}

\paragraph{Underlying CAD System} Our system is implemented as an add-on to Blender~\cite{blender}, a popular 3D modeling package. We support all Blender primitives ({\tt Cube}, {\tt Sphere}, etc.) and a number of common CAD operations, which can be composed arbitrarily. A full list of supported operations is included in supplemental materials. Our language is embedded in Python, and uses a subset of Python syntax, using custom interpretation of the Python AST to implement CAD operations and semantics for model parameters.

It may seem at first that our system could be implemented by naively applying an automatic differentiation tool, however there are several challenges that arise. Implementing on top of an existing CAD system implies that we interface with the underlying geometric kernel during interpretation in order to create and modify geometry. It is neither necessary nor desirable to differentiate kernel calls, nor to execute them during optimization. It is not possible to easily draw this distinction with off-the-shelf operator-overloading based automatic differentiation tools, which do not work across process boundaries and often rely on ~\cite{NEURIPS2019_PY, frostig2018compiling} dynamic language features to evaluate the function with custom datatypes that cannot be passed off to a CAD package. Fully differentiating a CAD kernel as done in~\cite{Banovic:2018, Banovic:2019} instead of providing a host language not only loses the safety guarantees offered by our DSL, but also incurs a 10-40x slowdown on model execution, which is untenable for interactive optimization.

As a result we explicitly construct our computation graph and compilation pass, instead of relying on an existing package to perform automatic differentiation. As mentioned in Section~\ref{sec:tracing}, our implementation still relies on the underlying CAD system to track mesh topology, and we index Blender vertices along with the vertex markers, updating those vertex coordinates directly with the results of evaluating $\F(P)$. As a result we completely avoid re-executing interpretation for changes that do not affect program structure.

\begin{figure*}
  \centering
  \includegraphics[width=\linewidth]{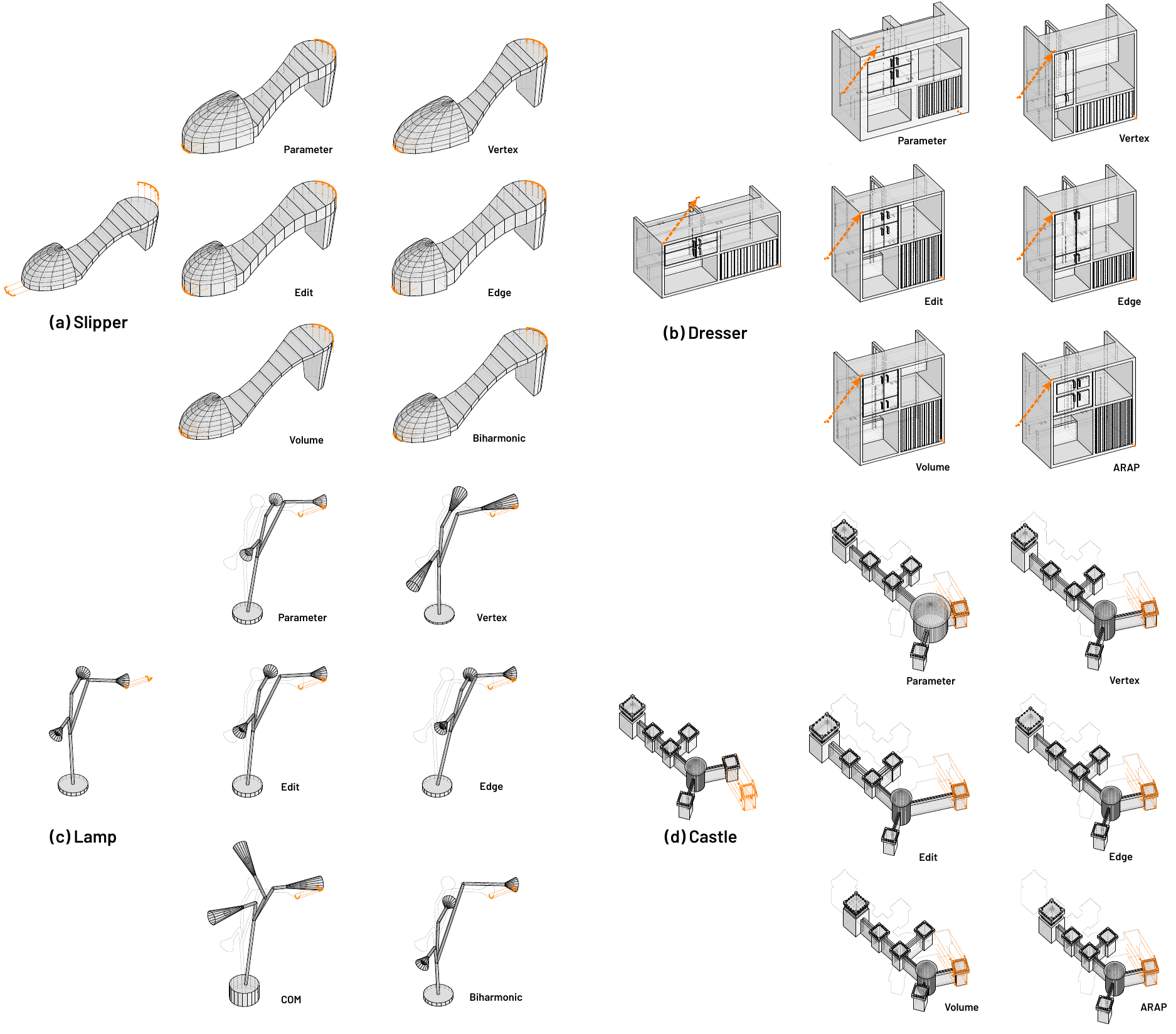}
  \caption{{\bf Option Gallery}. Each model is edited (orange), displaying the results our system produces by optimizing each objective. The edit specification is shown overlaid on each result to illustrate how closely the result matches the manipulation.}
  \label{fig:gallery}
\end{figure*}

\paragraph{Optimization and Performance} We minimize our objectives using the Sequential Least Squares Quadratic Programming (SLSQP) \cite{kraft1988software} optimizer through the SciPy library~\cite{2020SciPy-NMeth} with first order derivatives, primarily because of its ability to maintain constraints and make use of first-order information.Initial model interpretation is interactive, with models spanning hundreds of lines of code and over a thousand vertices usually interpreting in less than a second in our tests, running Python on a 2.5GHz Intel Core i7 Processor and 16 GB available RAM (Table~1). These models are comparable in size and complexity to complex CAD parts which are then combined into larger assemblies.

\section{Results}
\label{sec:results}

Our results focus on how bidirectional editing can be used to augment program editing through inverse synchronization steps. We present two key results with respect to our optimization procedure:
\begin{itemize}[leftmargin=*]
  \item When a manipulation is {\em ambiguous}, minimizing the objectives we have included tends to provide a diverse set of results. This increases the likelihood that one is close to the user's intent, and additionally serves to indicate which areas of the model must be further manipulated to remove ambiguity.
  \item When a manipulation is {\em unambiguous}, our objectives tend to converge on extremely similar results, approximating the closest point in the parameter space to that manipulation.
\end{itemize}
Together, these properties allow users to iterate quickly through a series of geometric edits, ultimately allowing complete control over model parameters through geometry, while maintaining the editing benefits of a program representation.

\begin{figure*}[ht]
  \centering
  \includegraphics[width=\linewidth]{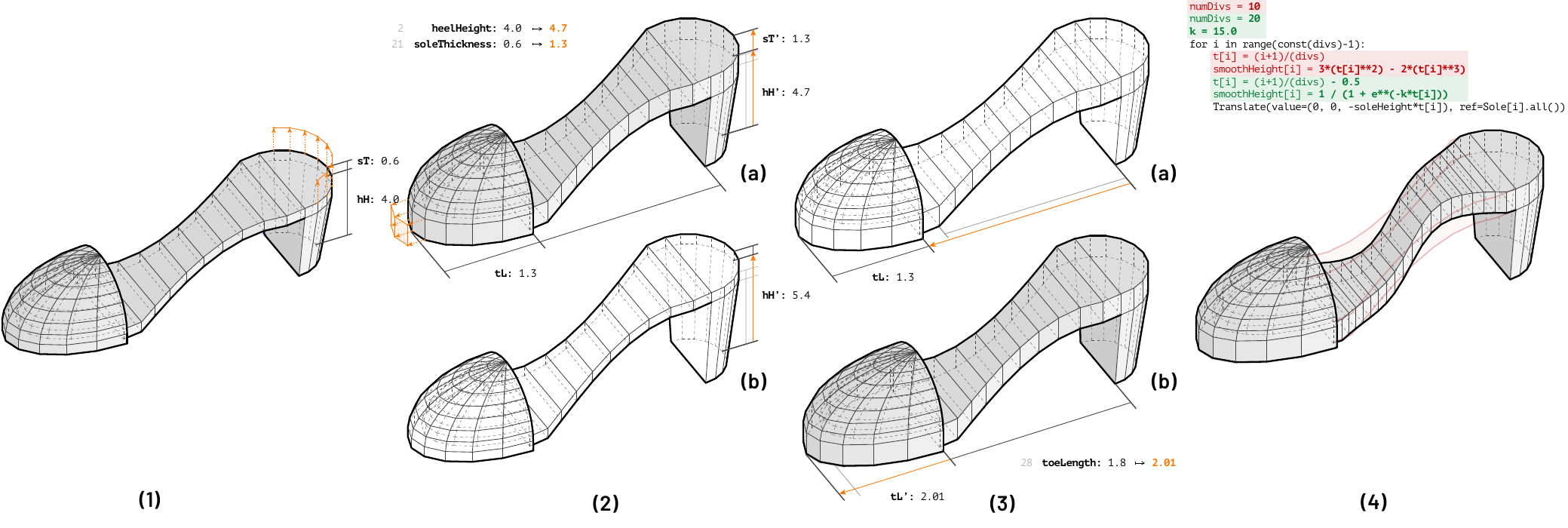}
  \caption{{\bf Slipper workflow.} A user iterates on a design for a freeform slipper. Vertices on the heel of the shoe are dragged up~(1) and an option is selected, and subsequently vertices are dragged outwards on the toe~(2a). After an option is selected for the toe edit~(3b), a program manipulation tweaking the sole curvature and mesh discretization is performed~(4).}
  \label{fig:slipper-workflow}
\end{figure*}

\subsection{Option Gallery}
Figure~\ref{fig:gallery} presents a model, an ambiguous manipulation, and the resulting diverse set of options returned by optimization of different objectives. Each objective can be described in terms of the geometric properties it preserves relative to the others.

The {\bf edit energy} tends to move all relevant parameters in the direction of the edit with no other optimization terms. Many parameters tend to change slightly, which can be frustrating when the user is trying to affect only a specific area of the model. We see this most clearly in the castle model, where the edit objective affects the castle (Figure~\ref{fig:gallery}d) far away from the edit and shifts it more than any other option.

Minimizing {\bf parameter change} can result in more drastic geometric changes (vertices moved, edges extended) than other options, though this is highly dependent on the program. Such changes can be seen clearly in the dresser and castle models; they are especially prevalent when the parameters are sensitive and small changes in value cause large geometric changes. This energy is most useful when a model directly exposes a subset of parameters that control the aspect the user wants to vary, or when the user intended a large manipulation that other objectives would penalize.

  {\bf Localized vertex-position} objectives can often end up deforming relative proportions of model elements (Lamp, Slipper) because they allow vertices close to the edit to move more than those further away.

As a result we see sharp variation close to the edit, while the rest of the model stays more or less fixed. This can also be seen in the dresser example, where most of the dressers' drawers retain their original height, except the top row of drawers, which stretch vertically in order to satisfy the objective. Moreover, in this example the left half of the model becomes much thinner horizontally than the right half -- partially because there are simply more vertices on the right half of the model, so moving them is penalized more.

  {\bf Localized edge-length preservation} in the dresser example shows a similar behavior. By minimizing the $L_1$ norm of edge length differences we are are encouraging sparsity, so the objective naturally prefers to stretch out the top-most vertical edges, as localization penalizes this change the least. In the slipper model, the edge objective prefers moving the toe forward to stretching it, because this results in overall less change in edge lengths, even when localizing the effects.

  {\bf Deformation objectives} do well at maintaining the overall geometric shape of the model, as they minimize the rate of local changes to the mesh imposed by satisfying the edit. This propagates changes smoothly from the edited region into the remainder of the model, resulting in a looser, but smoother, localization than the vertex and edge objectives (Slipper, Lamp).

Objectives preserving {\bf global properties}, such as Volume and Center of Mass, result in more non-local changes than the other objectives as they update parts of the model outside of the user's specified edit to maintain the overall property. This can be seen in the positions of un-manipulated arms in the lamp model, and the width of the bottom of the heel in the slipper model. Penalizing vertex movements prevents the changes from being extremely drastic.

\begin{figure*}
  \centering
  \includegraphics[width=\linewidth]{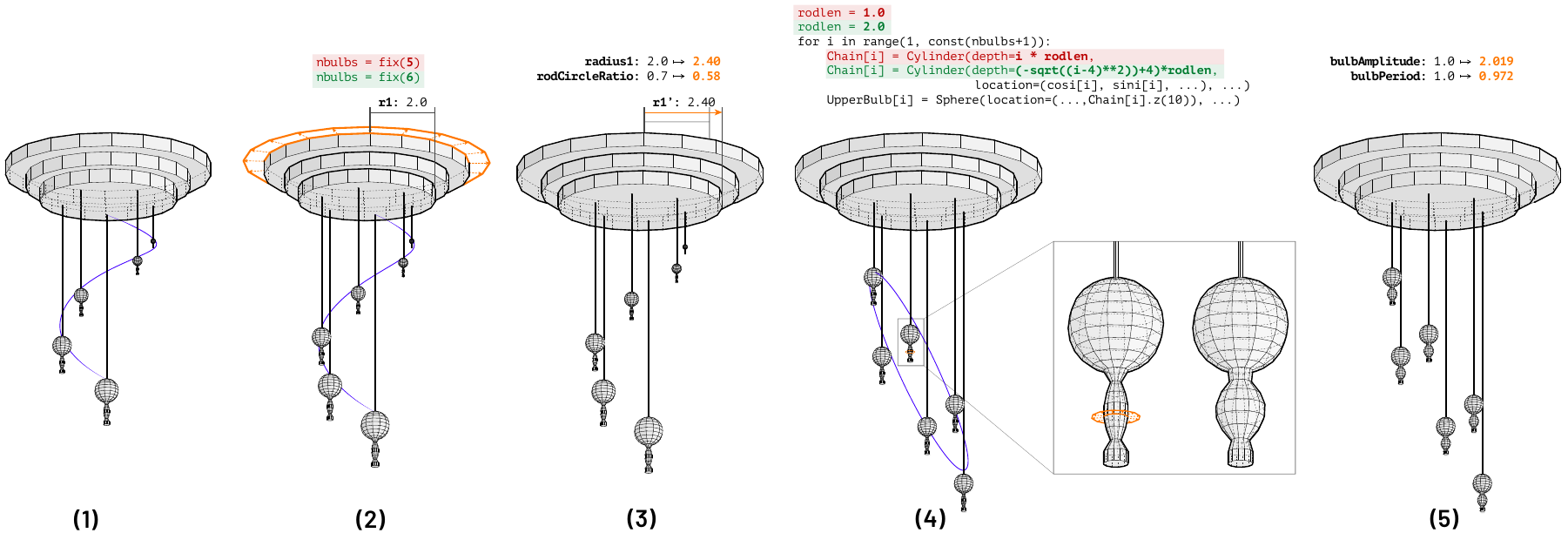}
  \caption{{\bf Chandelier workflow.} A user explores design variations for a chandelier model. She performs a program manipulation on the original, 5-bulb model~(1) to achieve a variant with 6~bulbs~(2), and scale out a ring of vertices around the top of the model to uniformly widen the base~(3). Another program change is performed to rearrange the bulb pattern~(4), and a final geometric edit widens a single bulb, resulting in all of the bulbs updating to match~(5).}
  \label{fig:chandelier-workflow}
\end{figure*}

Our bidirectional editing approach also addresses two common pain points in CAD modeling. Firstly, models encode constraints such as proportions, patterns, and bounds on the positions of elements. When editing CAD programs directly, users must reason about which parameters need to change together, and how to maintain explicitly-defined constraints. Our inverse editing step operates agnostically to how these constraints and patterns happen to be imposed in a particular program, finding a solution that preserves the encoded constraints regardless of how many parameters need to be adjusted. Patterns, such as the vertical stripes at the bottom right of the dresser (\ref{fig:gallery}b) or the crenelations on the castle towers (\ref{fig:gallery}d), maintain their structure through arbitrary parameter changes. Our method allows both painless editing of patterns and preservation of desirable invariants, both common challenges in CAD modeling.

Secondly, designed objects can have semantic properties that cannot be encoded purely in their geometry --- for example, a lamp must be stable and upright. It is also difficult to construct a program that limits the space of possible designs to exactly those that will be stable. Instead, optimizations such as Center of Mass (Figure~\ref{fig:gallery}c) can proxy the desired property and stay in the range of stable solutions without the user ever having to specify exactly how such limits are implemented.

\subsection{Slipper Workflow}

Figure~\ref{fig:slipper-workflow} illustrates how a user might edit a free-form slipper design to better match a specific target shape they have in mind. In the first manipulation~(1), the user wants to raise the heel and thicken the sole. To indicate this, she raises a few vertices on the heel of the shoe. But, the edit is ambiguous -- geometrically, the edit could be achieved by  either raising the heel, thickening the sole, or both; our objectives converge to two options that either increase a combination of both heel height and sole thickness~(2a); or raise the heel alone~(2b). Since the user desires to increase both of these aspects, they choose the first option.

The user then wants to extend the toe, and indicates this by dragging a few vertices out. Our system converges to two options that lengthen the sole~(3a) or lengthen the toe~(3b) respectively. Since the model happens to expose {\tt toeLength} explicitly as a parameter, the user chooses the option minimizing parameter energy~(3b) which matches their edit and focuses change on the {\tt toeLength} parameter, accomplishing their goal.

In the final step, our user wants to impose steeper curvature of the sole from toe to heel. To accomplish this she makes a program edit, changing the function controlling the curvature from a polynomial basis to a logistic one, and adjusts the number of divisions sampled to make the discretization finer.

\subsection{Chandelier Workflow}

Occasionally users do not have a specific goal in mind, but instead are attempting to explore possible design variations. This type of workflow is extremely common in digital asset design, where many similar but not identical model variations are desired.
Figure~\ref{fig:chandelier-workflow} demonstrates the steps a user might perform when manipulating a model of a chandelier to explore variations. After adjusting the number of bulbs by a program  edit~(1), the user adjusts the proportion of the radius of the chandelier base to the radius of the helical pattern, indicating it should be larger by uniformly scaling out a ring of vertices~(2). As they are not concerned with precise dimensions, they select the option generated by minimizing the biharmonic deformation energy~(3), which increases the radii of the bases without changing the helical radius by jointly adjusting two program parameters.

The user can then explore the space of variations of bulb arrangements by performing program edits, replacing the original helical arrangement with an elliptical one~(4). Each program option can be geometrically edited to test the output space of each bulb-arranging function. Finally, the user modifies a single bulb by widening a single loop around its lower half. The ARAP energy creates a smooth transition between this edit and the rest of the bulb. Moreover, updating the relevant parameters causes the remainder of the bulbs to update~(5). Automating this type of pattern change via direct manipulation is an advantage that arises only through bidirectional editing.

We note that for this example, previous work on editing man-made models fail to expose relationships between the radii of the three bases of the chandelier. In particular, iWires~\cite{Gal:2009:IWIRES} does not maintain a program representation and does not find any wires for the bulbs, which makes it impossible to achieve the last edit in the workflow, while \citet{bokeloh2012algebraic} does not support rotational patterns. Neither method can generate a chandelier with more or less bulbs while preserving the arrangement.

\subsection{Informal User Study}
We performed an initial informal study to examine how users interact with our tool. A few users with extensive background in CAD modeling across different domains (Engineering, Architecture, and Visual Arts) were introduced to our tool for 15 minutes. Users were then asked to a) match a 3D model to a target geometry overlaid in the interface; and b) manipulate a model to achieve a design variation, over three different models in 1.5 hours. We used a think-aloud protocol and asked participants to say what they were thinking as they worked with our tool. User testing highlighted the importance of promoting {\em diversity} in our results, as this allows users to understand how to adjust their manipulations even if the tool does not provide any correct result.

We observed that users initially perform several geometric edits when encountering a new model for the first time, intuiting a sense of how a model varies. When precise dimensions are not of concern, users quickly generate plausible design variations using geometric edits that change tens of parameters, and rarely consider more than two of the returned options.

Users were critical about difficulty in predicting how a model will change in response to an edit, and required a few test manipulations on any given model before it could be productively edited. Additionally, users occasionally encountered situations where the model program was constructed in a way that made an attempted edit infeasible. Users in this situation were usually uncertain what had occurred, and would benefit significantly from some form of automated explanation.

When browsing options generated in response to their edits, users usually only go through more than two alternatives if neither came close to satisfying their intent. Critically, if no alternatives correspond to user intent, users are almost instantly able to identify necessary changes to make to their edit to disambiguate by comparing the differences between results. Accordingly, users typically perform several iterations of geometric edits to achieve the target manipulations, changing tens of parameters in the process of doing so. All users found inverse editing with our tool significantly easier to use for creating design variations than the analogous CAD tools or program edits.

In conclusion, our method exposes a trade-off between giving users more options, and the time it takes to evaluate each option, versus the user identifying the ambiguity adjusting the manipulation to clarify intent. Interactive optimization performance is critical to bidirectional editing being useful, as users must be able to iterate quickly and adjust their manipulations.

\section{Limitations and Future Work}

The primary limitation of our approach is that while topological changes can be achieved by editing the program, our synchronization step by definition always preserves mesh topology; as it requires the function generating the mesh to be continuous, as well differentiable on most of its domain for the technique to work well. \citet{MB:2021:DAGA} illustrate another path forward where the gradient is not evaluated throughout the optimization and topological changes can be supported as long as UV coordinates are maintained. We see potential to merge such methods and accommodate a greater class of operators with higher performance, and note that the question of optimizing edits across topological change is unsolved in general.

Particularly, integer parameters such as loop bounds---which don't contribute to the gradient of a vertex and whose variation would create topologically different meshes---are always fixed in our system (Section~\ref{sec:lang}). Future work on optimizing through discrete changes would enable us to expand our solution to include arbitrary topological changes. Existing methods are beginning to allow differentiation through such discontinuities~\cite{BangaruMichel2021DiscontinuousAutodiff} via careful language design. One possible approach in this vein is to extend our current language to define references, user edits, and objective functions on collections of high-level elements of variable arity rather than explicitly over a fixed number of vertices.

We address ambiguity in a user manipulation by using heuristics in the form of different objective functions and energies to allow the user to manually disambiguate their intent if needed. This approach is not automatic, and when a given user intent is not within the program's output space, many objectives tend to converge to similar results while still differing from a user's intent. To address this, future work can use existing approaches to refining discrete parameter spaces, including design galleries ~\cite{marks1997design,shimizu2020design} and user-in-the loop optimization~\cite{koyama2020sequential}, in combination with program synthesis as a possible avenue for exploration. We suggest an approach where users can select an area of interest and a system can synthesize possible parameterizations that contain the desired output, which can then be explored using bidirectional editing.

Finally, an exciting avenue of future work revolves around the use of additional program information in geometry processing techniques. Geometry processing has traditionally been focused around polygonal meshes, and our method enables the augmentation of this mesh with a wealth of additional information garnered from analyzing the generating program -- we propose a sensitivity analysis of vertices with respect to parameters, methods for geometric constraint inference and maintenance, and program-aided design optimization integrated with the model editing process, to name a few. This would not only allow geometry processing techniques to be applied directly to CAD models, skipping often expensive translation steps, but also enable more powerful and versatile editing, analysis and optimization. % Programs <3 Geometry

\section{Conclusion}

3D CAD software is used by almost every designer in the world to create most of the virtual and man-made physical models we encounter. Moreover, direct editing of CAD programs is a challenge designers have been facing for multiple decades. This work argues that a combination of geometry, optimization, and programming languages methods can be intersected to devise new solutions that significantly reduce the existing burdens around CAD editing, allowing easier design iteration and customization. As an initial step, we described an implementation of {\em bidirectional} editing, where users leverage both programmatic expressivity as well as direct geometric manipulation. This approach realizes a significant advancement over the current state of the art of editing in CAD tools, ultimately allowing designers and engineers to work more quickly on a wide range of 3D design tasks.
\bibliographystyle{ACM-Reference-Format}
\bibliography{biblio.bib}


\begin{thebibliography}{52}

%%% the Web address.

\ifx \showCODEN    \undefined \def \showCODEN     #1{\unskip}     \fi
\ifx \showDOI      \undefined \def \showDOI       #1{#1}\fi
\ifx \showISBNx    \undefined \def \showISBNx     #1{\unskip}     \fi
\ifx \showISBNxiii \undefined \def \showISBNxiii  #1{\unskip}     \fi
\ifx \showISSN     \undefined \def \showISSN      #1{\unskip}     \fi
\ifx \showLCCN     \undefined \def \showLCCN      #1{\unskip}     \fi
\ifx \shownote     \undefined \def \shownote      #1{#1}          \fi
\ifx \showarticletitle \undefined \def \showarticletitle #1{#1}   \fi
\ifx \showURL      \undefined \def \showURL       {\relax}        \fi
% invisible to TeX
\providecommand\bibfield[2]{#2}
\providecommand\bibinfo[2]{#2}
\providecommand\natexlab[1]{#1}
\providecommand\showeprint[2][]{arXiv:#2}

\bibitem[\protect\citeauthoryear{Alba}{Alba}{2018}]%
        {Whatsth19:online}
\bibfield{author}{\bibinfo{person}{Michael Alba}.}
  \bibinfo{year}{2018}\natexlab{}.
\newblock \bibinfo{title}{What's the Difference Between Parametric and Direct
  Modeling?}
\newblock
  \bibinfo{howpublished}{\url{https://www.engineering.com/DesignSoftware/DesignSoftwareArticles/ArticleID/16587/Whats-the-Difference-Between-Parametric-and-Direct-Modeling.aspx}}.
\newblock
\newblock
\shownote{(Accessed on 09/06/2019).}


\bibitem[\protect\citeauthoryear{Bangaru, Michel, Mu, Bernstein, Li, and
  Ragan-Kelley}{Bangaru et~al\mbox{.}}{2021}]%
        {BangaruMichel2021DiscontinuousAutodiff}
\bibfield{author}{\bibinfo{person}{Sai Bangaru}, \bibinfo{person}{Jesse
  Michel}, \bibinfo{person}{Kevin Mu}, \bibinfo{person}{Gilbert Bernstein},
  \bibinfo{person}{Tzu-Mao Li}, {and} \bibinfo{person}{Jonathan Ragan-Kelley}.}
  \bibinfo{year}{2021}\natexlab{}.
\newblock \showarticletitle{Systematically Differentiating Parametric
  Discontinuities}.
\newblock \bibinfo{journal}{\emph{ACM Trans. Graph.}} \bibinfo{volume}{40},
  \bibinfo{number}{107} (\bibinfo{year}{2021}), \bibinfo{pages}{107:1--107:17}.
\newblock


\bibitem[\protect\citeauthoryear{Banovic}{Banovic}{2019}]%
        {Banovic:2019}
\bibfield{author}{\bibinfo{person}{Mladen Banovic}.}
  \bibinfo{year}{2019}\natexlab{}.
\newblock \showarticletitle{Efficient algorithmic differentiation of CAD
  frameworks}.
\newblock \bibinfo{journal}{\emph{Dissertation, Paderborn University}}
  (\bibinfo{year}{2019}).
\newblock


\bibitem[\protect\citeauthoryear{Banovic, Mykhaskiv, Auriemma, Walther,
  Legrand, and Muller}{Banovic et~al\mbox{.}}{2018}]%
        {Banovic:2018}
\bibfield{author}{\bibinfo{person}{Mladen Banovic}, \bibinfo{person}{Orest
  Mykhaskiv}, \bibinfo{person}{Salvatore Auriemma}, \bibinfo{person}{Andrea
  Walther}, \bibinfo{person}{Herve Legrand}, {and}
  \bibinfo{person}{Jens-Dominik Muller}.} \bibinfo{year}{2018}\natexlab{}.
\newblock \showarticletitle{Algorithmic differentiation of the Open CASCADE
  Technology CAD kernel and its coupling with an adjoint CFD solver}.
\newblock \bibinfo{journal}{\emph{Optimization Methods and Software}}
  \bibinfo{volume}{33}, \bibinfo{number}{4-6} (\bibinfo{year}{2018}),
  \bibinfo{pages}{813--828}.
\newblock
\urldef\tempurl%
\url{https://doi.org/10.1080/10556788.2018.1431235}
\showDOI{\tempurl}
\showeprint{https://doi.org/10.1080/10556788.2018.1431235}


\bibitem[\protect\citeauthoryear{Berndt, Fellner, and Havemann}{Berndt
  et~al\mbox{.}}{2005}]%
        {Berndt:2005:GML}
\bibfield{author}{\bibinfo{person}{Rene Berndt}, \bibinfo{person}{Dieter~W.
  Fellner}, {and} \bibinfo{person}{Sven Havemann}.}
  \bibinfo{year}{2005}\natexlab{}.
\newblock \showarticletitle{Generative 3D Models: A Key to More Information
  Within Less Bandwidth at Higher Quality}. In
  \bibinfo{booktitle}{\emph{Proceedings of the Tenth International Conference
  on 3D Web Technology}} \emph{(\bibinfo{series}{Web3D '05})}.
  \bibinfo{publisher}{ACM}, \bibinfo{pages}{111--121}.
\newblock


\bibitem[\protect\citeauthoryear{Bokeloh, Wand, Koltun, and Seidel}{Bokeloh
  et~al\mbox{.}}{2011}]%
        {bokeloh2011pattern}
\bibfield{author}{\bibinfo{person}{Martin Bokeloh}, \bibinfo{person}{Michael
  Wand}, \bibinfo{person}{Vladlen Koltun}, {and} \bibinfo{person}{Hans-Peter
  Seidel}.} \bibinfo{year}{2011}\natexlab{}.
\newblock \showarticletitle{Pattern-aware shape deformation using sliding
  dockers}. In \bibinfo{booktitle}{\emph{ACM Transactions on Graphics (TOG)}},
  Vol.~\bibinfo{volume}{30}. ACM, \bibinfo{pages}{123}.
\newblock


\bibitem[\protect\citeauthoryear{Bokeloh, Wand, Seidel, and Koltun}{Bokeloh
  et~al\mbox{.}}{2012}]%
        {bokeloh2012algebraic}
\bibfield{author}{\bibinfo{person}{Martin Bokeloh}, \bibinfo{person}{Michael
  Wand}, \bibinfo{person}{Hans-Peter Seidel}, {and} \bibinfo{person}{Vladlen
  Koltun}.} \bibinfo{year}{2012}\natexlab{}.
\newblock \showarticletitle{An algebraic model for parameterized shape
  editing}.
\newblock \bibinfo{journal}{\emph{ACM Transactions on Graphics (TOG)}}
  \bibinfo{volume}{31}, \bibinfo{number}{4} (\bibinfo{year}{2012}),
  \bibinfo{pages}{78}.
\newblock


\bibitem[\protect\citeauthoryear{Brunelli}{Brunelli}{2014}]%
        {Parametr28:online}
\bibfield{author}{\bibinfo{person}{Mark Brunelli}.}
  \bibinfo{year}{2014}\natexlab{}.
\newblock \bibinfo{title}{Parametric vs. Direct Modeling | PTC}.
\newblock
  \bibinfo{howpublished}{\url{https://www.ptc.com/en/cad-software-blog/parametric-vs-direct-modeling-which-side-are-you-on}}.
\newblock
\newblock
\shownote{(Accessed on 09/08/2019).}


\bibitem[\protect\citeauthoryear{Chugh, Hempel, Spradlin, and Albers}{Chugh
  et~al\mbox{.}}{2016}]%
        {Chugh:2016:1}
\bibfield{author}{\bibinfo{person}{Ravi Chugh}, \bibinfo{person}{Brian Hempel},
  \bibinfo{person}{Mitchell Spradlin}, {and} \bibinfo{person}{Jacob Albers}.}
  \bibinfo{year}{2016}\natexlab{}.
\newblock \showarticletitle{Programmatic and Direct Manipulation, Together at
  Last}. In \bibinfo{booktitle}{\emph{Proceedings of the 37th ACM SIGPLAN
  Conference on Programming Language Design and Implementation}}
  \emph{(\bibinfo{series}{PLDI '16})}. ACM, \bibinfo{address}{Santa Barbara,
  CA, USA}, \bibinfo{pages}{341--354}.
\newblock


\bibitem[\protect\citeauthoryear{Community}{Community}{2018}]%
        {blender}
\bibfield{author}{\bibinfo{person}{Blender~Online Community}.}
  \bibinfo{year}{2018}\natexlab{}.
\newblock \bibinfo{booktitle}{\emph{Blender - a 3D modelling and rendering
  package}}.
\newblock Blender Foundation, Stichting Blender Foundation, Amsterdam.
\newblock
\urldef\tempurl%
\url{http://www.blender.org}
\showURL{%
\tempurl}


\bibitem[\protect\citeauthoryear{Du, Inala, Pu, Spielberg, Schulz, Rus,
  Solar-Lezama, and Matusik}{Du et~al\mbox{.}}{2018}]%
        {du2018inversecsg}
\bibfield{author}{\bibinfo{person}{Tao Du}, \bibinfo{person}{Jeevana~Priya
  Inala}, \bibinfo{person}{Yewen Pu}, \bibinfo{person}{Andrew Spielberg},
  \bibinfo{person}{Adriana Schulz}, \bibinfo{person}{Daniela Rus},
  \bibinfo{person}{Armando Solar-Lezama}, {and} \bibinfo{person}{Wojciech
  Matusik}.} \bibinfo{year}{2018}\natexlab{}.
\newblock \showarticletitle{Inversecsg: Automatic conversion of 3d models to
  csg trees}.
\newblock \bibinfo{journal}{\emph{ACM Transactions on Graphics (TOG)}}
  \bibinfo{volume}{37}, \bibinfo{number}{6} (\bibinfo{year}{2018}),
  \bibinfo{pages}{1--16}.
\newblock


\bibitem[\protect\citeauthoryear{Frostig, Johnson, and Leary}{Frostig
  et~al\mbox{.}}{2018}]%
        {frostig2018compiling}
\bibfield{author}{\bibinfo{person}{Roy Frostig}, \bibinfo{person}{Matthew~James
  Johnson}, {and} \bibinfo{person}{Chris Leary}.}
  \bibinfo{year}{2018}\natexlab{}.
\newblock \showarticletitle{Compiling machine learning programs via high-level
  tracing}.
\newblock \bibinfo{journal}{\emph{Systems for Machine Learning}}
  (\bibinfo{year}{2018}).
\newblock


\bibitem[\protect\citeauthoryear{Gal, Sorkine, Mitra, and Cohen-Or}{Gal
  et~al\mbox{.}}{2009}]%
        {Gal:2009:IWIRES}
\bibfield{author}{\bibinfo{person}{Ran Gal}, \bibinfo{person}{Olga Sorkine},
  \bibinfo{person}{Niloy~J. Mitra}, {and} \bibinfo{person}{Daniel Cohen-Or}.}
  \bibinfo{year}{2009}\natexlab{}.
\newblock \showarticletitle{IWIRES: an analyze-and-edit approach to shape
  manipulation}.
\newblock \bibinfo{journal}{\emph{ACM Transactions on Graphics}}
  \bibinfo{volume}{28}, \bibinfo{number}{3} (\bibinfo{year}{2009}).
\newblock


\bibitem[\protect\citeauthoryear{Griewank and Walther}{Griewank and
  Walther}{2008}]%
        {griewank2008derivatives}
\bibfield{author}{\bibinfo{person}{A. Griewank} {and} \bibinfo{person}{A.
  Walther}.} \bibinfo{year}{2008}\natexlab{}.
\newblock \bibinfo{booktitle}{\emph{Evaluating Derivatives: Principles and
  Techniques of Algorithmic Differentiation, Second Edition}}.
\newblock \bibinfo{publisher}{Society for Industrial and Applied Mathematics
  (SIAM, 3600 Market Street, Floor 6, Philadelphia, PA 19104)}.
\newblock
\showISBNx{9780898717761}
\showLCCN{2008021064}
\urldef\tempurl%
\url{https://books.google.com/books?id=xoiiLaRxcbEC}
\showURL{%
\tempurl}


\bibitem[\protect\citeauthoryear{Guo, Jiang, Benes, Deussen, Zhang, Lischinski,
  and Huang}{Guo et~al\mbox{.}}{2020}]%
        {guo2020inverse}
\bibfield{author}{\bibinfo{person}{Jianwei Guo}, \bibinfo{person}{Haiyong
  Jiang}, \bibinfo{person}{Bedrich Benes}, \bibinfo{person}{Oliver Deussen},
  \bibinfo{person}{Xiaopeng Zhang}, \bibinfo{person}{Dani Lischinski}, {and}
  \bibinfo{person}{Hui Huang}.} \bibinfo{year}{2020}\natexlab{}.
\newblock \showarticletitle{Inverse Procedural Modeling of Branching Structures
  by Inferring L-Systems}.
\newblock \bibinfo{journal}{\emph{ACM Transactions on Graphics (TOG)}}
  \bibinfo{volume}{39}, \bibinfo{number}{5} (\bibinfo{year}{2020}),
  \bibinfo{pages}{1--13}.
\newblock


\bibitem[\protect\citeauthoryear{Hafner, Schumacher, Knoop, Auzinger, Bickel,
  and B{\"a}cher}{Hafner et~al\mbox{.}}{2019}]%
        {hafner2019x}
\bibfield{author}{\bibinfo{person}{Christian Hafner},
  \bibinfo{person}{Christian Schumacher}, \bibinfo{person}{Espen Knoop},
  \bibinfo{person}{Thomas Auzinger}, \bibinfo{person}{Bernd Bickel}, {and}
  \bibinfo{person}{Moritz B{\"a}cher}.} \bibinfo{year}{2019}\natexlab{}.
\newblock \showarticletitle{X-CAD: optimizing CAD models with extended finite
  elements}.
\newblock \bibinfo{journal}{\emph{ACM Transactions on Graphics (TOG)}}
  \bibinfo{volume}{38}, \bibinfo{number}{6} (\bibinfo{year}{2019}),
  \bibinfo{pages}{1--15}.
\newblock


\bibitem[\protect\citeauthoryear{Hempel, Lubin, and Chugh}{Hempel
  et~al\mbox{.}}{2019}]%
        {hempel2019sketch}
\bibfield{author}{\bibinfo{person}{Brian Hempel}, \bibinfo{person}{Justin
  Lubin}, {and} \bibinfo{person}{Ravi Chugh}.} \bibinfo{year}{2019}\natexlab{}.
\newblock \showarticletitle{Sketch-n-Sketch: Output-Directed Programming for
  SVG}. In \bibinfo{booktitle}{\emph{Proceedings of the 32nd Annual ACM
  Symposium on User Interface Software and Technology}}.
  \bibinfo{pages}{281--292}.
\newblock


\bibitem[\protect\citeauthoryear{Hottelier, Bod{\'{\i}}k, and Ryokai}{Hottelier
  et~al\mbox{.}}{2014}]%
        {Hottelier:14:PBM}
\bibfield{author}{\bibinfo{person}{Thibaud Hottelier}, \bibinfo{person}{Ras
  Bod{\'{\i}}k}, {and} \bibinfo{person}{Kimiko Ryokai}.}
  \bibinfo{year}{2014}\natexlab{}.
\newblock \showarticletitle{Programming by Manipulation for Layout}. In
  \bibinfo{booktitle}{\emph{The 27th Annual {ACM} Symposium on User Interface
  Software and Technology {UIST} '14}}. \bibinfo{pages}{231--241}.
\newblock


\bibitem[\protect\citeauthoryear{Hu, Schurr, Stevens, and Terwilliger}{Hu
  et~al\mbox{.}}{2011}]%
        {Hu:2011:BX}
\bibfield{author}{\bibinfo{person}{Zhenjiang Hu}, \bibinfo{person}{Andy
  Schurr}, \bibinfo{person}{Perdita Stevens}, {and} \bibinfo{person}{James~F.
  Terwilliger}.} \bibinfo{year}{2011}\natexlab{}.
\newblock \showarticletitle{Dagstuhl Seminar on Bidirectional Transformations
  (BX)}. In \bibinfo{booktitle}{\emph{Proceedings of the 2011 International
  Conference on Management of Data}} \emph{(\bibinfo{series}{SIGMOD '11})},
  Vol.~\bibinfo{volume}{30}. ACM, \bibinfo{address}{Athens, Greece},
  \bibinfo{pages}{35--39}.
\newblock


\bibitem[\protect\citeauthoryear{Jesus, Patow, Coelho, and Sousa}{Jesus
  et~al\mbox{.}}{2018}]%
        {jesus2018generalized}
\bibfield{author}{\bibinfo{person}{Diego Jesus}, \bibinfo{person}{Gustavo
  Patow}, \bibinfo{person}{Ant{\'o}nio Coelho}, {and}
  \bibinfo{person}{Antonio~Augusto Sousa}.} \bibinfo{year}{2018}\natexlab{}.
\newblock \showarticletitle{Generalized selections for direct control in
  procedural buildings}.
\newblock \bibinfo{journal}{\emph{Computers \& Graphics}}  \bibinfo{volume}{72}
  (\bibinfo{year}{2018}), \bibinfo{pages}{106--121}.
\newblock


\bibitem[\protect\citeauthoryear{Koyama, Sato, and Goto}{Koyama
  et~al\mbox{.}}{2020}]%
        {koyama2020sequential}
\bibfield{author}{\bibinfo{person}{Yuki Koyama}, \bibinfo{person}{Issei Sato},
  {and} \bibinfo{person}{Masataka Goto}.} \bibinfo{year}{2020}\natexlab{}.
\newblock \showarticletitle{Sequential gallery for interactive visual design
  optimization}.
\newblock \bibinfo{journal}{\emph{ACM Transactions on Graphics (TOG)}}
  \bibinfo{volume}{39}, \bibinfo{number}{4} (\bibinfo{year}{2020}),
  \bibinfo{pages}{88--1}.
\newblock


\bibitem[\protect\citeauthoryear{Kraevoy, Sheffer, Shamir, and
  Cohen-Or}{Kraevoy et~al\mbox{.}}{2008}]%
        {kraevoy2008non}
\bibfield{author}{\bibinfo{person}{Vladislav Kraevoy}, \bibinfo{person}{Alla
  Sheffer}, \bibinfo{person}{Ariel Shamir}, {and} \bibinfo{person}{Daniel
  Cohen-Or}.} \bibinfo{year}{2008}\natexlab{}.
\newblock \showarticletitle{Non-homogeneous resizing of complex models}. In
  \bibinfo{booktitle}{\emph{ACM Transactions on Graphics (TOG)}},
  Vol.~\bibinfo{volume}{27}. ACM, \bibinfo{pages}{111}.
\newblock


\bibitem[\protect\citeauthoryear{Kraft et~al\mbox{.}}{Kraft
  et~al\mbox{.}}{1988}]%
        {kraft1988software}
\bibfield{author}{\bibinfo{person}{Dieter Kraft} {et~al\mbox{.}}}
  \bibinfo{year}{1988}\natexlab{}.
\newblock \showarticletitle{A software package for sequential quadratic
  programming}.
\newblock  (\bibinfo{year}{1988}).
\newblock


\bibitem[\protect\citeauthoryear{Li, Luk{\'a}{\v{c}}, Gharbi, and
  Ragan-Kelley}{Li et~al\mbox{.}}{2020}]%
        {li2020differentiable}
\bibfield{author}{\bibinfo{person}{Tzu-Mao Li}, \bibinfo{person}{Michal
  Luk{\'a}{\v{c}}}, \bibinfo{person}{Micha{\"e}l Gharbi}, {and}
  \bibinfo{person}{Jonathan Ragan-Kelley}.} \bibinfo{year}{2020}\natexlab{}.
\newblock \showarticletitle{Differentiable vector graphics rasterization for
  editing and learning}.
\newblock \bibinfo{journal}{\emph{ACM Transactions on Graphics (TOG)}}
  \bibinfo{volume}{39}, \bibinfo{number}{6} (\bibinfo{year}{2020}),
  \bibinfo{pages}{1--15}.
\newblock


\bibitem[\protect\citeauthoryear{Lipp, Specht, Lau, Wonka, and Müller}{Lipp
  et~al\mbox{.}}{2019}]%
        {Lipp2019}
\bibfield{author}{\bibinfo{person}{M. Lipp}, \bibinfo{person}{M. Specht},
  \bibinfo{person}{C. Lau}, \bibinfo{person}{P. Wonka}, {and}
  \bibinfo{person}{P. Müller}.} \bibinfo{year}{2019}\natexlab{}.
\newblock \showarticletitle{Local Editing of Procedural Models}.
\newblock \bibinfo{journal}{\emph{Computer Graphics Forum}}
  \bibinfo{volume}{38} (\bibinfo{date}{05} \bibinfo{year}{2019}),
  \bibinfo{pages}{13--25}.
\newblock
\urldef\tempurl%
\url{https://doi.org/10.1111/cgf.13615}
\showDOI{\tempurl}


\bibitem[\protect\citeauthoryear{Lipp, Wonka, and Wimmer}{Lipp
  et~al\mbox{.}}{2008}]%
        {Lipp:Visual:2008}
\bibfield{author}{\bibinfo{person}{Markus Lipp}, \bibinfo{person}{Peter Wonka},
  {and} \bibinfo{person}{Michael Wimmer}.} \bibinfo{year}{2008}\natexlab{}.
\newblock \showarticletitle{Interactive Visual Editing of Grammars for
  Procedural Architecture}.
\newblock \bibinfo{journal}{\emph{ACM Transactions on Graphics}}
  \bibinfo{volume}{27}, \bibinfo{number}{3} (\bibinfo{date}{Aug.}
  \bibinfo{year}{2008}), \bibinfo{pages}{35:1--35:9}.
\newblock


\bibitem[\protect\citeauthoryear{Margossian}{Margossian}{2019}]%
        {margossian19autodiff}
\bibfield{author}{\bibinfo{person}{Charles~C. Margossian}.}
  \bibinfo{year}{2019}\natexlab{}.
\newblock \showarticletitle{A review of automatic differentiation and its
  efficient implementation}.
\newblock \bibinfo{journal}{\emph{WIREs Data Mining and Knowledge Discovery}}
  \bibinfo{volume}{9}, \bibinfo{number}{4} (\bibinfo{year}{2019}),
  \bibinfo{pages}{e1305}.
\newblock
\urldef\tempurl%
\url{https://doi.org/10.1002/widm.1305}
\showDOI{\tempurl}


\bibitem[\protect\citeauthoryear{Marks, Andalman, Beardsley, Freeman, Gibson,
  Hodgins, Kang, Mirtich, Pfister, Ruml, et~al\mbox{.}}{Marks
  et~al\mbox{.}}{1997}]%
        {marks1997design}
\bibfield{author}{\bibinfo{person}{Joe Marks}, \bibinfo{person}{Brad Andalman},
  \bibinfo{person}{Paul~A Beardsley}, \bibinfo{person}{William Freeman},
  \bibinfo{person}{Sarah Gibson}, \bibinfo{person}{Jessica Hodgins},
  \bibinfo{person}{Thomas Kang}, \bibinfo{person}{Brian Mirtich},
  \bibinfo{person}{Hanspeter Pfister}, \bibinfo{person}{Wheeler Ruml},
  {et~al\mbox{.}}} \bibinfo{year}{1997}\natexlab{}.
\newblock \showarticletitle{Design galleries: A general approach to setting
  parameters for computer graphics and animation}. In
  \bibinfo{booktitle}{\emph{Proceedings of the 24th annual conference on
  Computer graphics and interactive techniques}}. \bibinfo{pages}{389--400}.
\newblock


\bibitem[\protect\citeauthoryear{Michel and Boubekeur}{Michel and
  Boubekeur}{2021}]%
        {MB:2021:DAGA}
\bibfield{author}{\bibinfo{person}{Elie Michel} {and} \bibinfo{person}{Tamy
  Boubekeur}.} \bibinfo{year}{2021}\natexlab{}.
\newblock \showarticletitle{DAG Amendment for Inverse Control of Parametric
  Shapes}.
\newblock \bibinfo{journal}{\emph{ACM Transactions on Graphics}}
  \bibinfo{volume}{40}, \bibinfo{number}{4} (\bibinfo{year}{2021}),
  \bibinfo{pages}{173:1--173:14}.
\newblock


\bibitem[\protect\citeauthoryear{Mitra, Wand, Zhang, Cohen-Or, Kim, and
  Huang}{Mitra et~al\mbox{.}}{2014}]%
        {mitra2014structure}
\bibfield{author}{\bibinfo{person}{Niloy~J Mitra}, \bibinfo{person}{Michael
  Wand}, \bibinfo{person}{Hao Zhang}, \bibinfo{person}{Daniel Cohen-Or},
  \bibinfo{person}{Vladimir Kim}, {and} \bibinfo{person}{Qi-Xing Huang}.}
  \bibinfo{year}{2014}\natexlab{}.
\newblock \showarticletitle{Structure-aware shape processing}. In
  \bibinfo{booktitle}{\emph{ACM SIGGRAPH 2014 Courses}}. ACM,
  \bibinfo{pages}{13}.
\newblock


\bibitem[\protect\citeauthoryear{M\"{u}ller, Wonka, Haegler, Ulmer, and
  Van~Gool}{M\"{u}ller et~al\mbox{.}}{2006}]%
        {Muller:2006:Procedural}
\bibfield{author}{\bibinfo{person}{Pascal M\"{u}ller}, \bibinfo{person}{Peter
  Wonka}, \bibinfo{person}{Simon Haegler}, \bibinfo{person}{Andreas Ulmer},
  {and} \bibinfo{person}{Luc Van~Gool}.} \bibinfo{year}{2006}\natexlab{}.
\newblock \showarticletitle{Procedural Modeling of Buildings}.
\newblock \bibinfo{journal}{\emph{ACM Transactions on Graphics}}
  \bibinfo{volume}{25}, \bibinfo{number}{3} (\bibinfo{date}{July}
  \bibinfo{year}{2006}).
\newblock


\bibitem[\protect\citeauthoryear{Nandi, Wilcox, Panchekha, Blau, Grossman, and
  Tatlock}{Nandi et~al\mbox{.}}{2018}]%
        {nandi2018functional}
\bibfield{author}{\bibinfo{person}{Chandrakana Nandi}, \bibinfo{person}{James~R
  Wilcox}, \bibinfo{person}{Pavel Panchekha}, \bibinfo{person}{Taylor Blau},
  \bibinfo{person}{Dan Grossman}, {and} \bibinfo{person}{Zachary Tatlock}.}
  \bibinfo{year}{2018}\natexlab{}.
\newblock \showarticletitle{Functional programming for compiling and
  decompiling computer-aided design}.
\newblock \bibinfo{journal}{\emph{Proceedings of the ACM on Programming
  Languages}} \bibinfo{volume}{2}, \bibinfo{number}{ICFP}
  (\bibinfo{year}{2018}), \bibinfo{pages}{1--31}.
\newblock


\bibitem[\protect\citeauthoryear{Nandi, Willsey, Anderson, Wilcox, Darulova,
  Grossman, and Tatlock}{Nandi et~al\mbox{.}}{2020}]%
        {nandi2020synthesizing}
\bibfield{author}{\bibinfo{person}{Chandrakana Nandi}, \bibinfo{person}{Max
  Willsey}, \bibinfo{person}{Adam Anderson}, \bibinfo{person}{James~R Wilcox},
  \bibinfo{person}{Eva Darulova}, \bibinfo{person}{Dan Grossman}, {and}
  \bibinfo{person}{Zachary Tatlock}.} \bibinfo{year}{2020}\natexlab{}.
\newblock \showarticletitle{Synthesizing structured CAD models with equality
  saturation and inverse transformations}.
\newblock \bibinfo{journal}{\emph{vertex}} \bibinfo{volume}{9},
  \bibinfo{number}{15} (\bibinfo{year}{2020}).
\newblock


\bibitem[\protect\citeauthoryear{Nimier-David, Vicini, Zeltner, and
  Jakob}{Nimier-David et~al\mbox{.}}{2019}]%
        {nimier2019mitsuba}
\bibfield{author}{\bibinfo{person}{Merlin Nimier-David}, \bibinfo{person}{Delio
  Vicini}, \bibinfo{person}{Tizian Zeltner}, {and} \bibinfo{person}{Wenzel
  Jakob}.} \bibinfo{year}{2019}\natexlab{}.
\newblock \showarticletitle{Mitsuba 2: A retargetable forward and inverse
  renderer}.
\newblock \bibinfo{journal}{\emph{ACM Transactions on Graphics (TOG)}}
  \bibinfo{volume}{38}, \bibinfo{number}{6} (\bibinfo{year}{2019}),
  \bibinfo{pages}{1--17}.
\newblock


\bibitem[\protect\citeauthoryear{Paszke, Gross, Massa, Lerer, Bradbury, Chanan,
  Killeen, Lin, Gimelshein, Antiga, Desmaison, Kopf, Yang, DeVito, Raison,
  Tejani, Chilamkurthy, Steiner, Fang, Bai, and Chintala}{Paszke
  et~al\mbox{.}}{2019}]%
        {NEURIPS2019_PY}
\bibfield{author}{\bibinfo{person}{Adam Paszke}, \bibinfo{person}{Sam Gross},
  \bibinfo{person}{Francisco Massa}, \bibinfo{person}{Adam Lerer},
  \bibinfo{person}{James Bradbury}, \bibinfo{person}{Gregory Chanan},
  \bibinfo{person}{Trevor Killeen}, \bibinfo{person}{Zeming Lin},
  \bibinfo{person}{Natalia Gimelshein}, \bibinfo{person}{Luca Antiga},
  \bibinfo{person}{Alban Desmaison}, \bibinfo{person}{Andreas Kopf},
  \bibinfo{person}{Edward Yang}, \bibinfo{person}{Zachary DeVito},
  \bibinfo{person}{Martin Raison}, \bibinfo{person}{Alykhan Tejani},
  \bibinfo{person}{Sasank Chilamkurthy}, \bibinfo{person}{Benoit Steiner},
  \bibinfo{person}{Lu Fang}, \bibinfo{person}{Junjie Bai}, {and}
  \bibinfo{person}{Soumith Chintala}.} \bibinfo{year}{2019}\natexlab{}.
\newblock \showarticletitle{PyTorch: An Imperative Style, High-Performance Deep
  Learning Library}.
\newblock In \bibinfo{booktitle}{\emph{Advances in Neural Information
  Processing Systems 32}}, \bibfield{editor}{\bibinfo{person}{H.~Wallach},
  \bibinfo{person}{H.~Larochelle}, \bibinfo{person}{A.~Beygelzimer},
  \bibinfo{person}{F.~d\textquotesingle Alch\'{e}-Buc},
  \bibinfo{person}{E.~Fox}, {and} \bibinfo{person}{R.~Garnett}} (Eds.).
  \bibinfo{publisher}{Curran Associates, Inc.}, \bibinfo{pages}{8024--8035}.
\newblock
\urldef\tempurl%
\url{http://papers.neurips.cc/paper/9015-pytorch-an-imperative-style-high-performance-deep-learning-library.pdf}
\showURL{%
\tempurl}


\bibitem[\protect\citeauthoryear{Reddy, Guerrero, Fisher, Li, and Mitra}{Reddy
  et~al\mbox{.}}{2020}]%
        {reddy2020discovering}
\bibfield{author}{\bibinfo{person}{Pradyumna Reddy}, \bibinfo{person}{Paul
  Guerrero}, \bibinfo{person}{Matt Fisher}, \bibinfo{person}{Wilmot Li}, {and}
  \bibinfo{person}{Niloy~J Mitra}.} \bibinfo{year}{2020}\natexlab{}.
\newblock \showarticletitle{Discovering pattern structure using differentiable
  compositing}.
\newblock \bibinfo{journal}{\emph{ACM Transactions on Graphics (TOG)}}
  \bibinfo{volume}{39}, \bibinfo{number}{6} (\bibinfo{year}{2020}),
  \bibinfo{pages}{1--15}.
\newblock


\bibitem[\protect\citeauthoryear{Schulz, Shamir, Levin, Sitthi-Amorn, and
  Matusik}{Schulz et~al\mbox{.}}{2014}]%
        {schulz2014design}
\bibfield{author}{\bibinfo{person}{Adriana Schulz}, \bibinfo{person}{Ariel
  Shamir}, \bibinfo{person}{David~IW Levin}, \bibinfo{person}{Pitchaya
  Sitthi-Amorn}, {and} \bibinfo{person}{Wojciech Matusik}.}
  \bibinfo{year}{2014}\natexlab{}.
\newblock \showarticletitle{Design and fabrication by example}.
\newblock \bibinfo{journal}{\emph{ACM Transactions on Graphics (TOG)}}
  \bibinfo{volume}{33}, \bibinfo{number}{4} (\bibinfo{year}{2014}),
  \bibinfo{pages}{62}.
\newblock


\bibitem[\protect\citeauthoryear{Schulz, Xu, Zhu, Zheng, Grinspun, and
  Matusik}{Schulz et~al\mbox{.}}{2017}]%
        {Schulz:2017}
\bibfield{author}{\bibinfo{person}{Adriana Schulz}, \bibinfo{person}{Jie Xu},
  \bibinfo{person}{Bo Zhu}, \bibinfo{person}{Changxi Zheng},
  \bibinfo{person}{Eitan Grinspun}, {and} \bibinfo{person}{Wojciech Matusik}.}
  \bibinfo{year}{2017}\natexlab{}.
\newblock \showarticletitle{Interactive Design Space Exploration and
  Optimization for CAD Models}.
\newblock \bibinfo{journal}{\emph{ACM Transactions on Graphics}}
  \bibinfo{volume}{36}, \bibinfo{number}{4} (\bibinfo{date}{7}
  \bibinfo{year}{2017}).
\newblock


\bibitem[\protect\citeauthoryear{Shimizu, Fisher, Paris, McCann, and
  Fatahalian}{Shimizu et~al\mbox{.}}{2020}]%
        {shimizu2020design}
\bibfield{author}{\bibinfo{person}{Evan Shimizu}, \bibinfo{person}{Matthew
  Fisher}, \bibinfo{person}{Sylvain Paris}, \bibinfo{person}{James McCann},
  {and} \bibinfo{person}{Kayvon Fatahalian}.} \bibinfo{year}{2020}\natexlab{}.
\newblock \showarticletitle{Design Adjectives: A Framework for Interactive
  Model-Guided Exploration of Parameterized Design Spaces}. In
  \bibinfo{booktitle}{\emph{Proceedings of the 33rd Annual ACM Symposium on
  User Interface Software and Technology}}. \bibinfo{pages}{261--278}.
\newblock


\bibitem[\protect\citeauthoryear{Siemens}{Siemens}{2017}]%
        {synctechinfo}
\bibfield{author}{\bibinfo{person}{Siemens}.} \bibinfo{year}{2017}\natexlab{}.
\newblock \bibinfo{title}{Synchronous Technology - Going beyond traditional
  modeling approaches to solve design challenges}.
\newblock
\newblock
\urldef\tempurl%
\url{https://www.plm.automation.siemens.com/media/global/en/Siemens-PLM-Synchronous-technology-going-beyond-mi-63435_tcm27-28553.pdf}
\showURL{%
\tempurl}


\bibitem[\protect\citeauthoryear{Smelik, Tutenel, Bidarra, and Benes}{Smelik
  et~al\mbox{.}}{2014}]%
        {smelik2014survey}
\bibfield{author}{\bibinfo{person}{Ruben~M Smelik}, \bibinfo{person}{Tim
  Tutenel}, \bibinfo{person}{Rafael Bidarra}, {and} \bibinfo{person}{Bedrich
  Benes}.} \bibinfo{year}{2014}\natexlab{}.
\newblock \showarticletitle{A survey on procedural modelling for virtual
  worlds}. In \bibinfo{booktitle}{\emph{Computer Graphics Forum}},
  Vol.~\bibinfo{volume}{33}. Wiley Online Library, \bibinfo{pages}{31--50}.
\newblock


\bibitem[\protect\citeauthoryear{Sorkine and Alexa}{Sorkine and Alexa}{2007}]%
        {sorkine2007rigid}
\bibfield{author}{\bibinfo{person}{Olga Sorkine} {and} \bibinfo{person}{Marc
  Alexa}.} \bibinfo{year}{2007}\natexlab{}.
\newblock \showarticletitle{As-rigid-as-possible surface modeling}. In
  \bibinfo{booktitle}{\emph{Symposium on Geometry processing}},
  Vol.~\bibinfo{volume}{4}. \bibinfo{pages}{109--116}.
\newblock


\bibitem[\protect\citeauthoryear{Sorkine, Cohen-Or, Lipman, Alexa, R{\"o}ssl,
  and Seidel}{Sorkine et~al\mbox{.}}{2004}]%
        {sorkine2004laplacian}
\bibfield{author}{\bibinfo{person}{Olga Sorkine}, \bibinfo{person}{Daniel
  Cohen-Or}, \bibinfo{person}{Yaron Lipman}, \bibinfo{person}{Marc Alexa},
  \bibinfo{person}{Christian R{\"o}ssl}, {and} \bibinfo{person}{H-P Seidel}.}
  \bibinfo{year}{2004}\natexlab{}.
\newblock \showarticletitle{Laplacian surface editing}. In
  \bibinfo{booktitle}{\emph{Proceedings of the 2004 Eurographics/ACM SIGGRAPH
  symposium on Geometry processing}}. \bibinfo{pages}{175--184}.
\newblock


\bibitem[\protect\citeauthoryear{Strater}{Strater}{2016}]%
        {SolvedRe10:online}
\bibfield{author}{\bibinfo{person}{Jeff Strater}.}
  \bibinfo{year}{2016}\natexlab{}.
\newblock \bibinfo{title}{Solved: Re: Direct vs History based modelling - Page
  2 - Autodesk Community- Fusion 360}.
\newblock
  \bibinfo{howpublished}{\url{https://forums.autodesk.com/t5/fusion-360-design-validate/direct-vs-history-based-modelling/m-p/6717750\#M84483}}.
\newblock
\newblock
\shownote{(Accessed on 09/06/2019).}


\bibitem[\protect\citeauthoryear{Talton, Lou, Lesser, Duke, M{\v{e}}ch, and
  Koltun}{Talton et~al\mbox{.}}{2011}]%
        {talton2011metropolis}
\bibfield{author}{\bibinfo{person}{Jerry~O Talton}, \bibinfo{person}{Yu Lou},
  \bibinfo{person}{Steve Lesser}, \bibinfo{person}{Jared Duke},
  \bibinfo{person}{Radom{\'\i}r M{\v{e}}ch}, {and} \bibinfo{person}{Vladlen
  Koltun}.} \bibinfo{year}{2011}\natexlab{}.
\newblock \showarticletitle{Metropolis procedural modeling}.
\newblock \bibinfo{journal}{\emph{ACM Transactions on Graphics (TOG)}}
  \bibinfo{volume}{30}, \bibinfo{number}{2} (\bibinfo{year}{2011}),
  \bibinfo{pages}{11}.
\newblock


\bibitem[\protect\citeauthoryear{TensorFlow}{TensorFlow}{2020}]%
        {tfXLA}
\bibfield{author}{\bibinfo{person}{TensorFlow}.}
  \bibinfo{year}{2020}\natexlab{}.
\newblock \bibinfo{title}{XLA Architecture}.
\newblock
  \bibinfo{howpublished}{\url{https://www.tensorflow.org/xla/architecture}}.
\newblock
\newblock
\shownote{(Accessed on 01/26/2021).}


\bibitem[\protect\citeauthoryear{Vanegas, Garcia-Dorado, Aliaga, Benes, and
  Waddell}{Vanegas et~al\mbox{.}}{2012}]%
        {vanegas2012inverse}
\bibfield{author}{\bibinfo{person}{Carlos~A Vanegas}, \bibinfo{person}{Ignacio
  Garcia-Dorado}, \bibinfo{person}{Daniel~G Aliaga}, \bibinfo{person}{Bedrich
  Benes}, {and} \bibinfo{person}{Paul Waddell}.}
  \bibinfo{year}{2012}\natexlab{}.
\newblock \showarticletitle{Inverse design of urban procedural models}.
\newblock \bibinfo{journal}{\emph{ACM Transactions on Graphics (TOG)}}
  \bibinfo{volume}{31}, \bibinfo{number}{6} (\bibinfo{year}{2012}),
  \bibinfo{pages}{168}.
\newblock


\bibitem[\protect\citeauthoryear{Virtanen, Gommers, Oliphant, Haberland, Reddy,
  Cournapeau, Burovski, Peterson, Weckesser, Bright, {van der Walt}, Brett,
  Wilson, Millman, Mayorov, Nelson, Jones, Kern, Larson, Carey, Polat, Feng,
  Moore, {VanderPlas}, Laxalde, Perktold, Cimrman, Henriksen, Quintero, Harris,
  Archibald, Ribeiro, Pedregosa, {van Mulbregt}, and {SciPy 1.0
  Contributors}}{Virtanen et~al\mbox{.}}{2020}]%
        {2020SciPy-NMeth}
\bibfield{author}{\bibinfo{person}{Pauli Virtanen}, \bibinfo{person}{Ralf
  Gommers}, \bibinfo{person}{Travis~E. Oliphant}, \bibinfo{person}{Matt
  Haberland}, \bibinfo{person}{Tyler Reddy}, \bibinfo{person}{David
  Cournapeau}, \bibinfo{person}{Evgeni Burovski}, \bibinfo{person}{Pearu
  Peterson}, \bibinfo{person}{Warren Weckesser}, \bibinfo{person}{Jonathan
  Bright}, \bibinfo{person}{St{\'e}fan~J. {van der Walt}},
  \bibinfo{person}{Matthew Brett}, \bibinfo{person}{Joshua Wilson},
  \bibinfo{person}{K.~Jarrod Millman}, \bibinfo{person}{Nikolay Mayorov},
  \bibinfo{person}{Andrew R.~J. Nelson}, \bibinfo{person}{Eric Jones},
  \bibinfo{person}{Robert Kern}, \bibinfo{person}{Eric Larson},
  \bibinfo{person}{C~J Carey}, \bibinfo{person}{{\.I}lhan Polat},
  \bibinfo{person}{Yu Feng}, \bibinfo{person}{Eric~W. Moore},
  \bibinfo{person}{Jake {VanderPlas}}, \bibinfo{person}{Denis Laxalde},
  \bibinfo{person}{Josef Perktold}, \bibinfo{person}{Robert Cimrman},
  \bibinfo{person}{Ian Henriksen}, \bibinfo{person}{E.~A. Quintero},
  \bibinfo{person}{Charles~R. Harris}, \bibinfo{person}{Anne~M. Archibald},
  \bibinfo{person}{Ant{\^o}nio~H. Ribeiro}, \bibinfo{person}{Fabian Pedregosa},
  \bibinfo{person}{Paul {van Mulbregt}}, {and} \bibinfo{person}{{SciPy 1.0
  Contributors}}.} \bibinfo{year}{2020}\natexlab{}.
\newblock \showarticletitle{{{SciPy} 1.0: Fundamental Algorithms for Scientific
  Computing in Python}}.
\newblock \bibinfo{journal}{\emph{Nature Methods}}  \bibinfo{volume}{17}
  (\bibinfo{year}{2020}), \bibinfo{pages}{261--272}.
\newblock
\urldef\tempurl%
\url{https://doi.org/10.1038/s41592-019-0686-2}
\showDOI{\tempurl}


\bibitem[\protect\citeauthoryear{Willis, Pu, Luo, Chu, Du, Lambourne,
  Solar-Lezama, and Matusik}{Willis et~al\mbox{.}}{2020}]%
        {willis2020fusion}
\bibfield{author}{\bibinfo{person}{Karl~DD Willis}, \bibinfo{person}{Yewen Pu},
  \bibinfo{person}{Jieliang Luo}, \bibinfo{person}{Hang Chu},
  \bibinfo{person}{Tao Du}, \bibinfo{person}{Joseph~G Lambourne},
  \bibinfo{person}{Armando Solar-Lezama}, {and} \bibinfo{person}{Wojciech
  Matusik}.} \bibinfo{year}{2020}\natexlab{}.
\newblock \showarticletitle{Fusion 360 Gallery: A Dataset and Environment for
  Programmatic CAD Reconstruction}.
\newblock \bibinfo{journal}{\emph{arXiv preprint arXiv:2010.02392}}
  (\bibinfo{year}{2020}).
\newblock


\bibitem[\protect\citeauthoryear{Wonka, Wimmer, Sillion, and Ribarsky}{Wonka
  et~al\mbox{.}}{2003}]%
        {Wonka:2003:Architecture}
\bibfield{author}{\bibinfo{person}{Peter Wonka}, \bibinfo{person}{Michael
  Wimmer}, \bibinfo{person}{Fran\c{c}ois Sillion}, {and}
  \bibinfo{person}{William Ribarsky}.} \bibinfo{year}{2003}\natexlab{}.
\newblock \showarticletitle{Instant Architecture}.
\newblock \bibinfo{journal}{\emph{ACM Transactions on Graphics}}
  \bibinfo{volume}{22}, \bibinfo{number}{3} (\bibinfo{date}{July}
  \bibinfo{year}{2003}), \bibinfo{pages}{669--677}.
\newblock


\bibitem[\protect\citeauthoryear{Yares}{Yares}{2013}]%
        {Thefaile61:online}
\bibfield{author}{\bibinfo{person}{Evan Yares}.}
  \bibinfo{year}{2013}\natexlab{}.
\newblock \bibinfo{title}{The failed promise of parametric CAD part 1: From the
  beginning}.
\newblock
  \bibinfo{howpublished}{\url{https://www.3dcadworld.com/the-failed-promise-of-parametric-cad/}}.
\newblock
\newblock
\shownote{(Accessed on 09/06/2019).}


\bibitem[\protect\citeauthoryear{Zheng, Fu, Cohen-Or, Kin-Chung~Au, and
  Tai}{Zheng et~al\mbox{.}}{2011}]%
        {Zheng:2011:Controllers}
\bibfield{author}{\bibinfo{person}{Youyi Zheng}, \bibinfo{person}{Hongbo Fu},
  \bibinfo{person}{Daniel Cohen-Or}, \bibinfo{person}{Oscar Kin-Chung~Au},
  {and} \bibinfo{person}{Chiew-Lan Tai}.} \bibinfo{year}{2011}\natexlab{}.
\newblock \showarticletitle{Component-wise Controllers for Structure-Preserving
  Shape Manipulation}.
\newblock \bibinfo{journal}{\emph{Computer Graphics Forum}}
  \bibinfo{volume}{30}, \bibinfo{number}{2} (\bibinfo{year}{2011}),
  \bibinfo{pages}{563--572}.
\newblock


\end{thebibliography}
\appendix
\section{Derivations}
\label{app:1}

We describe how we compute the gradients $\frac{\partial E_{bh}(P)}{\partial p_i}$ and $\frac{\partial E_{ARAP}(P)}{\partial p_i}$, for a parameter $p_i {\in} P$.
Let $E(P)=E_{bh}(P)=tr(D(P)^TQD(P)$. Then,

\[
  \begin{aligned} \displaystyle &
                \frac{\partial E(P)}{\partial p_i}=\left(QD(P)+Q^TD(P)\right) \frac {\partial \left(D(P)^TQD(P)\right)}{\partial p_i} = \\&
                \left(QD(P)+Q^TD(P)\right)(Q+Q^T)D(P) \frac {\partial D(P)}{\partial p_i}
  \end{aligned}
\]

Recall that $D(P)=\F(P){-}\F(P_0)$, and we compute $\frac {\partial D(P)}{\partial p_i} = \frac {\partial \F(P)}{\partial p_i}$ using forward-mode automatic differentiation.

Similarly, let $E(P)=E_{ARAP}(P)$.
We use the same alternating minimization strategy as in~\cite{sorkine2007rigid}: for a given fixed set of rigid transformations~$\{R_i\}$, we find parameters~$P'$ that minimize~$E(P')$. Then, we find the rigid transformations~$\{R_i\}$ that minimize~$E(P)$ for the given set of parameters~$P$.
$\{R_i\}$ is computed using SVD (Equation (6)~\cite{sorkine2007rigid}) of the covariance matrix $S_i$ (Equation (5)~\cite{sorkine2007rigid}). Then, $P'$ is computed by solving Equation~(9)~\cite{sorkine2007rigid}, with the only difference that we compute $\frac{\partial E(P')}{\partial p_i'}=\frac{\partial E(P')}{\partial V_i(P')}\frac{\partial V_i(P')}{\partial p_i'}$.

\end{document}